\documentstyle[12pt]{article}

  \def\pp{{\mathchoice
             {%displaystyle
              \kern 1pt%
              \raise 1pt
              \vbox{\hrule width5pt height0.4pt depth0pt
                    \kern -2pt
                    \hbox{\kern 2.3pt
                          \vrule width0.4pt height6pt depth0pt
                          }
                    \kern -2pt
                    \hrule width5pt height0.4pt depth0pt}%
                    \kern 1pt
           }
            {%textstyle
              \kern 1pt%
              \raise 1pt
              \vbox{\hrule width4.3pt height0.4pt depth0pt
                    \kern -1.8pt
                    \hbox{\kern 1.95pt
                          \vrule width0.4pt height5.4pt depth0pt
                          }
                    \kern -1.8pt
                    \hrule width4.3pt height0.4pt depth0pt}%
                    \kern 1pt
            }
            {%scriptstyle
              \kern 0.5pt%
              \raise 1pt
              \vbox{\hrule width4.0pt height0.3pt depth0pt
                    \kern -1.9pt  %[e]=0.15pt
                    \hbox{\kern 1.85pt
                          \vrule width0.3pt height5.7pt depth0pt
                          }
                    \kern -1.9pt
                    \hrule width4.0pt height0.3pt depth0pt}%
                    \kern 0.5pt
            }
            {%scriptscriptstyle
              \kern 0.5pt%
              \raise 1pt
              \vbox{\hrule width3.6pt height0.3pt depth0pt
                    \kern -1.5pt
                    \hbox{\kern 1.65pt
                          \vrule width0.3pt height4.5pt depth0pt
                          }
                    \kern -1.5pt
                    \hrule width3.6pt height0.3pt depth0pt}%
                    \kern 0.5pt%}
            }
        }}

  \def\mm{{\mathchoice
                       {%displaystyle
                             \kern 1pt
               \raise 1pt    \vbox{\hrule width5pt height0.4pt depth0pt
                                  \kern 2pt
                                  \hrule width5pt height0.4pt depth0pt}
                             \kern 1pt}
                       {%textstyle
                            \kern 1pt
               \raise 1pt \vbox{\hrule width4.3pt height0.4pt depth0pt
                                  \kern 1.8pt
                                  \hrule width4.3pt height0.4pt depth0pt}
                             \kern 1pt}
                       {%scriptstyle
                            \kern 0.5pt
               \raise 1pt
                            \vbox{\hrule width4.0pt height0.3pt depth0pt
                                  \kern 1.9pt
                                  \hrule width4.0pt height0.3pt depth0pt}
                            \kern 1pt}
                       {%scriptscriptstyle
                           \kern 0.5pt
             \raise 1pt  \vbox{\hrule width3.6pt height0.3pt depth0pt
                                  \kern 1.5pt
                                  \hrule width3.6pt height0.3pt depth0pt}
                           \kern 0.5pt}
                       }}

\catcode`@=11
\def\un#1{\relax\ifmmode\@@underline#1\else
        $\@@underline{\hbox{#1}}$\relax\fi}
\catcode`@=12

\def\a{\alpha}
\def\b{\beta}

\def\d{\delta}

\def\f{\phi}
\def\g{\gamma}

\def\j{\psi}

\def\x{\xi}

\def\G{\Gamma}

\def\ve{\varepsilon}

\def\bo{{\hbox{\large$\Box$}}}                 % D'Alembertian
\def\Bo{\raise -1pt\hbox{\bo{\hskip 0.03in}}}  % improved D'Alembertian

\def\pa{\partial}                                       % curly d
\def\pr{\prod}                                          % product
\def\TH{{\raise.2ex\hbox{$\displaystyle \bigodot$}\mskip-4.7mu \llap H \;}}
\def\face{{\raise.2ex\hbox{$\displaystyle \bigodot$}\mskip-2.2mu \llap {$\ddot
        \smile$}}}                                      % happy face
\def\dg{\sp\dagger}                                     % hermitian conjugate
\def\sp#1{{}^{#1}}                              % superscript (unaligned)
\def\Hat#1{\widehat{#1}}                        % big hat
\def\Bar#1{\overline{#1}}                       % big bar
\def\leftrightarrowfill{$\mathsurround=0pt \mathord\leftarrow \mkern-6mu
        \cleaders\hbox{$\mkern-2mu \mathord- \mkern-2mu$}\hfill
        \mkern-6mu \mathord\rightarrow$}
\def\dvec#1{\vbox{\ialign{##\crcr
        \leftrightarrowfill\crcr\noalign{\kern-1pt\nointerlineskip}
        $\hfil\displaystyle{#1}\hfil$\crcr}}}           % <--> accent
\def\frac#1#2{{\textstyle{#1\over\vphantom2\smash{\raise.20ex
        \hbox{$\scriptstyle{#2}$}}}}}                   % fraction
\def\sfrac#1#2{{\vphantom1\smash{\lower.5ex\hbox{\small$#1$}}\over
        \vphantom1\smash{\raise.4ex\hbox{\small$#2$}}}} % alternate fraction
\def\bfrac#1#2{{\vphantom1\smash{\lower.5ex\hbox{$#1$}}\over
        \vphantom1\smash{\raise.3ex\hbox{$#2$}}}}       % "
\def\afrac#1#2{{\vphantom1\smash{\lower.5ex\hbox{$#1$}}\over#2}}    % "
\def\[{\lfloor{\hskip 0.35pt}\!\!\!\lceil}  % improved commutator left bracket
\def\]{\rfloor{\hskip 0.35pt}\!\!\!\rceil} % improved commutator right bracket
\def\ud#1#2{^{#1}{}_{#2}}

\def\fracm#1#2{\hbox{\large{${\frac{{#1}}{{#2}}}$}}}
\def\fracmm#1#2{{{#1}\over{#2}}}                        % fractions
\def\half{{\fracm12}}
\def\ha{\half}
\def\fracmm#1#2{{{#1}\over{#2}}}

\def\tr{{\rm tr}}                                    % traces

\def\ul{\underline}                         % underline

\def\un{{\underline n}}

\def\low#1{{\raise -3pt\hbox{${\hskip 0.75pt}\!_{#1}$}}}  % big indices
\def\Low#1{{\raise -8pt\hbox{$\!\!\!\!\!\!_{#1}$}}~\;}
\def\low#1{{\raise -3pt\hbox{${\hskip 1.0pt}\!_{#1}$}}}

\def\Hat#1{\widehat{#1}}

\def\plpl{{+\!\!\!\!\!{\hskip 0.009in}{\raise -1.0pt\hbox{$_+$}}
{\hskip 0.0008in}}}

\def\mimi{{-\!\!\!\!\!{\hskip 0.009in}{\raise -1.0pt\hbox{$_-$}}
{\hskip 0.0008in}}}

\def\-{{\hskip 1.5pt}\hbox{-}}

\newskip\humongous \humongous=0pt plus 1000pt minus 1000pt
\def\caja{\mathsurround=0pt}
\def\eqalign#1{\,\vcenter{\openup2\jot \caja
        \ialign{\strut \hfil$\displaystyle{##}$&$
        \displaystyle{{}##}$\hfil\crcr#1\crcr}}\,}
\newif\ifdtup

\def\ref#1{$\sp{#1)}$}

\def\pl#1#2#3{Phys.~Lett.~{\bf {#1}B} (19{#2}) #3}
\def\np#1#2#3{Nucl.~Phys.~{\bf B{#1}} (19{#2}) #3}
\def\prl#1#2#3{Phys.~Rev.~Lett.~{\bf #1} (19{#2}) #3}
\def\pr#1#2#3{Phys.~Rev.~{\bf D{#1}} (19{#2}) #3}
\def\cqg#1#2#3{Class.~and Quantum Grav.~{\bf {#1}} (19{#2}) #3}
\def\cmp#1#2#3{Commun.~Math.~Phys.~{\bf {#1}} (19{#2}) #3}

\def\mpl#1#2#3{Mod.~Phys.~Lett.~{\bf A{#1}} (19{#2}) #3}

\def\ibid#1#2#3{{\it ibid.}~{\bf {#1}} (19{#2}) #3}

\parskip=\medskipamount           % space between paragraphs (LaTeX)
\thicklines                     % thick straight lines for pictures (LaTeX)

\begin{document}

% =========================== title page =================================

\thispagestyle{empty}             % no heading or foot on title page (LaTeX)

\def\border{                                            % UH + PSU border
        \setlength{\unitlength}{1mm}
        \newcount\xco
        \newcount\yco
        \xco=-24
        \yco=12
        \begin{picture}(140,0)
        \put(-20,11){\tiny Institut f\"ur Theoretische Physik Universit\"at
Hannover~~ Institut f\"ur Theoretische Physik Universit\"at Hannover~~
Institut f\"ur Theoretische Physik Hannover}
        \put(-20,-241.5){\tiny Department of Physics Penn State University~ 
Department of Physics Penn State University~ Department of Physics Penn 
State University~ Department of Physics}
        \end{picture}
        \par\vskip-8mm}

\def\headpic{                                           % UH heading
        \indent
        \setlength{\unitlength}{.8mm}
        \thinlines
        \par
        \begin{picture}(29,16)
        \put(5,16){\line(1,0){4}}
        \put(80,16){\line(1,0){4}}
        \put(85,16){\line(1,0){4}}
        \put(92,16){\line(1,0){4}}

        \put(85,0){\line(1,0){4}}
        \put(89,8){\line(1,0){3}}
        \put(92,0){\line(1,0){4}}

        \put(85,0){\line(0,1){16}}
        \put(96,0){\line(0,1){16}}
        \put(79,0){\line(0,1){16}}
        \put(80,0){\line(0,1){16}}
        \put(89,0){\line(0,1){16}}
        \put(92,0){\line(0,1){16}}
        \put(79,16){\oval(8,32)[bl]}
        \put(80,16){\oval(8,32)[br]}

        \end{picture}
        \par\vskip-6.5mm
        \thicklines}

% Optional !

\noindent 
DESY 95 -- 255 \hfill December 1995  \\
ITP--UH -- 31/95 \hfill hep-th/9601072  \\
PSU ~95 -- 166 \hfill ISSN 0418 -- 9833 \\ 

\begin{center}
{\Large\bf 

          SEVEN--SPHERE AND THE  EXCEPTIONAL 
              N=7 AND N=8 SUPERCONFORMAL 
                       ALGEBRAS                     }
\\
\vglue.2in
Murat G\"unaydin

{\it Department of Physics, Penn State University, University Park,
PA 16802, USA}\\
{\sl murat@phys.psu.edu} \\
\vglue.1in
and\\
\vglue.1in
Sergei V. Ketov~\footnote{Supported in part by the `Deutsche 
Forschungsgemeinschaft' and the `Volkswagen Stiftung'} ${}^{,}$ 
\footnote{On leave of absence from:
High Current Electronics Institute of the Russian Academy of Sciences,
\newline ${~~~~~}$ Siberian Branch, Akademichesky~4, Tomsk 634055, Russia}

{\it Institut f\"ur Theoretische Physik, Universit\"at Hannover}\\
{\it Appelstra\ss{}e 2, 30167 Hannover, Germany}\\
{\sl ketov@itp.uni-hannover.de}
\end{center}

\begin{center}
{\Large\bf Abstract}
\end{center}

We study realizations of the exceptional non-linear (quadratically generated, 
or $W$-type) $N=8$ and $N=7$ superconformal algebras with $Spin(7)$ and $G_2$
affine  symmetry currents, respectively. Both the $N=8$ and $N=7$ algebras 
admit unitary highest-weight representations in terms of a single boson and free 
fermions in ${\underline{8}}$ of $Spin(7)$ and ${\underline{7}}$ of $G_2$, with 
the central charges $c_8=26/5$ and $c_7=5$, respectively. Furthermore, we show 
that the general coset Ans\"{a}tze for the $N=8$ and $N=7$ algebras naturally 
lead to the coset spaces $SO(8)\times U(1)/SO(7)$ and $SO(7)\times U(1)/G_2\,$,
respectively, as the additional consistent solutions for certain values of the
central charge. The coset space $SO(8)/SO(7)$ is  the seven-sphere $S^7$, 
whereas the space $SO(7)/G_2$ represents the seven-sphere with torsion, 
$S^7_{\rm T}$. The division algebra of octonions and the associated triality 
properties of $SO(8)$ play an essential role in all these realizations. We 
also comment on some possible applications of our results to string theory.

% ======================= END of TITLE PAGE =============================

\newpage
\baselineskip=25pt

% =======================================================================

\section{Introduction}

Infinite conformal symmetry in two dimensions is the fundamental underlying
symmetry of string theory \cite{gsw}, and it plays an essential role in the 
understanding of the critical behaviour of two-dimensional physical systems. 
Similarly, supersymmetric extensions of the infinite-dimensional conformal 
algebra underlie various superstring theories. For example,
 space-time supersymmetric 
classical vacua of superstring theories in various dimensions are described by
extended superconformal field theories. Extended superconformal symmetry also 
has applications to integrable systems and to topological field theories as 
well \cite{jb,book}.

The finite-dimensional (global) subgroup of the two-dimensional conformal 
group $SO(2,2)$ is not simple, and it decomposes as $SO(2,2)\simeq SO(2,1) 
\times SO(2,1)$, with the two $SO(2,1)$ factors acting on left and right 
movers, respectively. This allows one to have different number of 
supersymmetries in the left and right moving sectors. A complete 
classification of supersymmetric extensions of the {\it finite}-dimensional 
global conformal group in two dimensions 
was given in \cite{gst}. However, not all  finite-dimensional (global)
superconformal algebras admit extensions to {\it infinite}-dimensional linear 
superconformal algebras with generators of non-negative conformal 
dimensions. The maximal number $(N)$ of supersymmetries such linear infinite 
superconformal algebras can have is four \cite{rs,af,sstp,htt}. 

It is possible to have supersymmetric extensions of the Virasoro algebra with 
$N>4$, while retaining the requirement that the generators have non-negative 
conformal dimensions. However, the price one has to pay is either to have a 
non-linear superconformal algebra, as in the Bershadsky-Knizhnik algebras 
\cite{kn,be}, or introduce field-dependent structure constants, as in the 
so-called {\it soft\/} $N=8$ algebra introduced in ref.~\cite{estps} and 
further studied in refs.~\cite{nb,bcp,cp,sp}. The  soft $N=8$ algebra appears
as the algebra of first-class constraints in the Green-Schwarz superstring 
action in ten dimensions~\cite{nb}.  Such field-dependent `structure 
constants' also appear in the symmetry algebras of the two-dimensional {\it 
locally} supersymmetric non-linear sigma-models (NLSMs) with $N>4$, where they
may even become non-chiral due to non-trivial mixing between the left- and 
right-moving modes via dilaton couplings, when the number of supersymmetries 
exceeds eight \cite{nw}. The Grassmannian symmetric spaces 
$$\fracmm{SO(8,m)}{SO(8)\times SO(m)}\qquad\quad m\geq 1,
\eqno(1.1) $$
 appear as solutions for the $N=8$ locally supersymmetric NLSM target 
manifolds \cite{bsn,wtn}. The soft algebras usually have no restrictions on 
their central extensions, while their `structure constants' are, in fact, 
functions on the target manifold.

The $N$-extended superconformal algebras of the type introduced by Bershadsky 
and Knizhnik \cite{kn,be} comprise generators of conformal dimension $2$, $3/2$
 and $1$ only. They contain (i) the Virasoro subalgebra, (ii) $\,N$ real 
supercurrents of conformal dimension $3/2$, whose operator products give the  
stress tensor of dimension $2$, symmetry currents of dimension $1$, and
terms that are quadratic in the symmetry currents, (iii) satisfy the Jacobi
 `identities', and (iv) have the usual spin-statistics relation.  Under the 
requirement of reductivity, a complete classification of such algebras was 
given in refs.~\cite{fl,bo}. Being ``reductive'' means that they linearise in 
the limit when their central charges go to infinity. In this limit, the 
infinite-dimensional vacuum-preserving algebra becomes a finite superalgebra
containing the finite (global) conformal algebra. Thus, the full classification
 of such non-linear superconformal algebras \cite{fl,bo} follows from the 
classification of finite-dimensional (global) superconformal algebras given in
ref.~\cite{gst}. There are three infinite classical families (for either the 
right- or the left-moving modes) , 
$$osp(N|2;{\bf R})~,\quad su(1,1|N)~,\quad osp(4^*|2N)~,\eqno(1.2)$$
a one-parameter family of the $N=4$ algebras, and two exceptional 
superconformal algebras with $N=7$ and $N=8$ supersymmetries. 

All the extended superconformal algebras can be viewed as arising from the 
quantum hamiltonian (Drinfeld-Sokolov-type) reduction of affine Lie 
superalgebras \cite{imp}. In particular, the vacuum-preserving subalgebras of 
the $N=7$ and $N=8$ exceptional superconformal algebras in the limit of 
infinite central charge are the exceptional finite Lie superalgebras $G(3)$ 
and $F(4)$ (in the Ka\v{c} notation \cite{kac}).~\footnote{See  
refs.~\cite{fk,snr,dvn} for details about the $G(3)$ and $F(4)$. Another real 
form of $F(4)$ corresponds to $N=2$ superconformal symmetry in five space-time 
dimensions \cite{mg1}.} Hence, it is not surprising that the quantum 
hamiltonian reduction of the affine Lie superalgebras $\Hat{G(3)}$ and 
$\Hat{F(4)}$ just yields the N=7 and N=8 superconformal algebras, respectively
 \cite{imp}. The orthogonal and unitary series of eq.~(1.2) are 
often referred to as the Bershadsky-Knizhnik superconformal 
algebras~\cite{kn,be}. The non-linear $N=4$ superconformal algebras were 
first obtained from the linear $N=4$ superconformal algebra by factoring 
out four free fermions and one boson \cite{gs,gptv}. The infinite classical 
family of non-linear superconformal algebras corresponding to $su(1,1|N)$ for 
$N>2$ does not admit unitary representations of the highest-weight 
type~\cite{kn,schoutens}. Similarly, the non-linear superconformal algebras 
corresponding to the symplectic series $osp(4^*|2N)$ do not admit unitary 
representations for $N>1$ either~\cite{bo}. The BRST operators of the
Bershadsky-Knizhnik-type superconformal algebras were studied in
refs.~\cite{ssvn,es,ke3}.

The main purpose of our investigation in this paper is to study possible coset
space realizations for the $N=7$ and $N=8$ exceptional superconformal algebras.
Thereby we will also answer the question as to whether or not there exist 
rational (unitary) superconformal field theories with the `exceptional'\/ $N=7$
or $N=8$ supersymmetry. When using the method of quantum hamiltonian 
reduction, the standard Wakimoto construction known for any affine Lie 
(super)algebra \cite{jb,book} allows one to obtain a free field 
(Feigin-Fuchs) representation for any extended non-linear superconformal 
algebra \cite{imp}. Though being practical for a calculation of the screening 
operators as well as the correlation functions in superconformal field theory,
using this method for constructing unitary highest-weight irreducible 
representations requires some additional techniques. For example, one still
needs to find zeroes of the Ka\v{c} determinant associated with a given 
(Verma) module and its null (singular) vectors, which may be a hard problem 
for the extended superconformal algebras at large $N$. On the other hand,
the coset construction can, in principle, answer the question of existence of 
rational superconformal field theories in a relatively simple and 
straightforward way. Therefore, we shall assume  that all the conformal fields 
in our construction come from the gauged (1,0) supersymmetric 
{\it Wess-Zumino-Novikov-Witten} (WZNW) models. 

Many of the results about unitary highest-weight representations of the 
{\it linear\/} $N=2$ and $N=4$ superconformal algebras were obtained in the 
past by using known results concerning superconformal algebras with lower $N$.
In the non-linear case, this method is obviously of a limited use, since the 
naive tensoring of representations is no longer valid. Thus we shall use the
coset space method directly, in studying the unitary highest-weight 
representations of the exceptional non-linear superconformal algebras. 
The coset construction 
is well-known to be a powerful tool in the two-dimensional (super)conformal 
field theory \cite{gko}, and it is presumably able to deliver all rational 
theories (modulo a permutation of fusion rules, or making an orbifold from the
coset by modding it with respect to a dicrete not-free-acting 
symmetry~\cite{jb,book}). The generalizations of the coset space method are 
known for the $N=2$ extended supersymmetry \cite{ks}, as well as for the $N=4$
supersymmetry \cite{vp,st,mg93,gk}. Though the coset space methods were 
invented to study representations of the linear extended superconformal 
algebras, they have also been extended to the non-linear $N=4$ superconformal 
algebras as well \cite{vp,st,mg93,gk}.

Our paper is organized as follows. In sect.~2, we provide the necessary 
algebraical and group-theoretical background in seven and eight dimensions, 
which simultaneously introduces our notation. In sect.~3, we review the 
non-linear $N=7$ and $N=8$ superconformal algebras, using the language of the
{\it operator product expansions} (OPEs). Our main results about the 
$N=8$ and $N=7$ coset constructions are presented in sect.~4.
 Our conclusions are summarized in sect.~5. The two appendices provide
 relevant identities for the octonionic structure constants (Appendix A), and 
some details about the supersymmetry part of the $N=8$ exceptional 
superconformal algebra (Appendix B).
\vglue.2in

\section{A review of the properties of octonions, their automorphism group and 
gamma matrices in seven dimensions}

In this section we shall review  some known results about the division algebra
 of octonions, its automorphism group $G_2$, and gamma matrices in seven 
dimensions. These results will be used in later sections, in our study of the 
exceptional superconformal algebras.
\vglue.2in

\subsection{Division algebra of octonions}
Many of the special properties of various mathematical structures in seven and
eight dimensions are related to the octonions.  The eight-dimensional 
division algebra of octonions {\bf  O} is one of the four division algebras 
that exist over the real numbers. An arbitrary octonion $q$ can be 
expanded as \cite{gg}
$$ q =\sum^7_{a=0} q_a\hat{e}_a~,\quad {\rm all}~~q_a~~{\rm are~~real~~
numbers}~,\eqno(2.1)$$
where $\hat{e}_0=1$ represents the identity element, and the imaginary  
octonion units  $\hat{e}_m$, $m=1,2,\ldots,7$, satisfy the multiplication rule
$$\hat{e}_m\hat{e}_n = -\d_{mn} + C_{mnp}\hat{e}_p~.\eqno(2.2)$$
with $C_{mnp}$ being the totally antisymmetric structure constants. The seven 
imaginary units close under commutation. However, they do not form a Lie 
algebra under commutation due to the non-associativity of octonions. The 
`Jacobian' of three elements is given by ~\footnote{We do not
 distinguish between co- and contra-variant indices. All (anti)symmetrizations
are defined with unit weight.}
$$ \[ \hat{e}_m, \[ \hat{e}_n , \hat{e}_p \] \] + {\rm cyclic~~permutations} = 
 3 C_{mnpq} \hat{e}_q~,\eqno(2.3) $$
where the non-vanishing totally antisymmetric tensor $C_{mnpq}$ is defined as
$$ C^k_{~[mn}C_{p]kq}=C_{mnpq}\neq 0~.\eqno(2.4)$$
The tensor $C_{mnpq}$ is dual to the tensor $C_{mnp}$ in seven dimensions:
$$C_{mnpq} = \fracmm{1}{6} \ve_{mnpqrst} C_{rst}~.\eqno(2.5) $$

We use the basis given in ref.~\cite{gg}, for which  the constants $C_{mnp}$ 
read as:
$$ C_{123}=C_{147}=C_{165}=C_{246}=C_{257}=C_{354}=C_{367}=1~,\eqno(2.6)$$
while all  the other non-vanishing components are determined by the total
antisymmetry. Given eq.~(2.6), the non-vanishing (equal to one) components of 
$C_{mnpq}$ are given by the following values of $(mnpq)$ \cite{snr}:
$$ (1,2,7,6)~,\quad  (1,2,4,5)~,\quad  (1,3,4,6)~,\quad  (1,3,5,7)~,$$ 
$$ (2,3,7,4)~,\quad  (2,3,5,6)~,\quad {\rm and}\quad (4,5,6,7)~,\eqno(2.7)$$
with the rest being fixed by the total antisymmetry.

The identities satisfied by the tensors  $C^{mnp}$ and $C^{mnpq}$ have been 
extensively studied in the literature~\cite{gg,snr,cdfn,wn,dgt,gn}).
 Among them, one has
$$\{ C^p,C^q \}_{mn} \equiv C^p_{~mk} C^q_{~kn}+ C^q_{~mk} C^p_{~kn}=
\d^p_m\d^q_n + \d^p_n\d^q_m - 2\d^{pq}\d_{mn}~, \eqno(2.8)$$
and  
$$ C_{mnpq}= -C^k_{~mn}C^k_{~pq} -\d_{mq}\d_{np} + \d_{mp}\d_{nq}~.\eqno(2.9)$$
A list of other useful identities, extending those found in 
refs.~\cite{gg,cdfn,wn,dgt,gn}, is given in Appendix A.
\vglue.2in

\subsection{$G_2\,$,~ $Spin(7)$~, ~$SO(8)$~, and Octonions}

The automorphism group of octonions is the exceptional group $G_2$. The
automorphisms together with left and right multiplications by unit octonions
generate the group $SO(8)$ \cite{gg}. The operation of simultaneous
 multiplication from the left by a unit octonion $q$ and right multiplication
by the octonion conjugate $\bar{q}$ together with the automorphisms
generate the group $SO(7)$ \cite{gg}. 
  
The  $8 \times 8$ gamma matrices  $\g^i_{ab}$  in seven dimensions, which
satisfy the Clifford algebra:
$$\{ \g^i,\g^j \} =2\d^{ij}{\bf 1}_8~,\eqno(2.10)$$
where $i,j,\ldots=1,2,\ldots,7$, and $a,b,\ldots=1,2,\ldots,8$, 
 can be written in terms of the octonionic structure
constants \cite{gg,cdfn,wn,dgt}. First, let's trivially extend $C^i_{jk}$ to 
$C^i_{ab}$
by setting $C^i_{ab}=C^i_{jk}$ whenever $a(=j)$ and $b(=k)$ are not equal to 
$8$, while defining $C^i_{ab}$ to be zero whenever $a$ or $b$ is equal to $8$.
The hermitian (purely imaginary and antisymmetric) gamma matrices in seven 
dimensions can then be chosen   as  
$$ \g^i_{ab} = i\left( C^i_{~ab} \pm \d_{ia}\d_{b8} \mp \d_{ib}\d_{a8}\right)~,
\eqno(2.11)$$
where the signs are correlated. Both options for the signs in eq.~(2.11) will
be exploited in sect.~4. In later sections we shall use the notation $\g^i_{ab}$ for the upper 
sign choice, whereas the notation $\tilde{\g}^i_{ab}$ is going to be used for 
the lower sign choice, in order to avoid confusion.

The antisymmetric products of gamma matrices are defined as usual, with
unit weight, {\it viz.}
$$ \g^{ij\cdots k}=\g^{[i}\g^{j}\cdots \g^{k]}~.\eqno(2.12)$$

 The  antisymmetric self-dual and antiself-dual tensors 
$C^{\pm}_{IJKL}$,  
($I,J,\ldots = 1,2,\ldots, 8 $) in eight dimensions will be defined as in 
refs.~\cite{cdfn,wn,dgt}:
$$C^{\pm}_{ijkl}=C_{ijkl}~,\quad {\rm and}\quad C^{\pm}_{ijk8}=\pm C_{ijk}~,
\eqno(2.13)$$
With the above
choices of gamma matrices one finds 
$$\eqalign{
\g^{ij}_{ab}~&~ = C_{ijab} + \d^i_a\d^j_b-\d^i_b\d^j_a \pm C^{~ij}_a\d_{b8}
\mp  C^{~ij}_b\d_{a8}\cr
{} ~&~ = C^{\pm}_{ijab} +  \d^i_a\d^j_b-\d^i_b\d^j_a~.\cr}\eqno(2.14)$$

A bit more effort is needed to calculate  $\g^{ijkl}$, and we summarize some
of the details below. First, it is straightforward to verify that
$$\eqalign{
\g^{ij}_{ac}\g_{cb}^{kl} = ~&~ -C^{\pm}_{ijac}C^{\pm}_{klbc}+
 C^{\pm}_{ijak}\d_{lb} -  C^{\pm}_{ijal}\d_{kb} +  C^{\pm}_{kljb}\d_{ia} \cr
{} ~&~ -  C^{\pm}_{klib}\d_{ja} +\d_{ia}\d_{jk}\d_{lb}-\d_{ia}\d_{jl}\d_{kb}-
\d_{ik}\d_{ja}\d_{lb} + \d_{il}\d_{ja}\d_{kb}~,\cr}$$
and, hence,
$$\eqalign{
\g^{[ij}_{ac}\g_{cb}^{kl]} = ~&~ -C^{\pm~[ij}_{ac}C^{\pm~kl]}_{bc} 
- 2C^{\pm}_{[ijk}{}^a\d_{l]}^b - 2C^{\pm}_{[ijk}{}^b\d_{l]}^a \cr
{} ~&~ =  -  C^{\pm~[ij}_{ap}C^{\pm~kl]}_{bp} -  C^{~[ij}_{a}C^{~kl]}_b 
- 2C^{\pm}_{[ijk}{}^a\d_{l]}^b - 2C^{\pm}_{[ijk}{}^b\d_{l]}^a~, \cr}$$
where we have used eq.~(A.3), in particular. Taking now $a=m$ and $b=n$ yields
$$\g^{[ij}_{mc}\g^{kl]}_{cn}=\d_{mn}C^{ijkl} + 4C_{m[ijk}\d_{l]n} + 
4C_{n[ijk}\d_{l]m}~,$$
For $a=b=8$ one finds
$$\g^{[ij}_{8c}\g^{kl]}_{c8}=-C^{~[ij}_pC^{~kl]}_p=C^{ijkl}~,$$
and taking $a=m$, $b=8$ yields
$$\g^{[ij}_{mc}\g^{kl]}_{c8}=-C^{[ij}_{~~mp}C^{~kl]}_p-2C^{[ijk}\d^{l]}_m=
-4C^{[ijk}\d^{l]}_m~.$$
Thus we find 
$$\g^{ijkl}_{ab} = \d_{ab}C^{ijkl} + 4C^{\pm}_{a[ijk}\d_{l]b} +
 4C^{\pm}_{b[ijk}\d_{l]a}~,\eqno(2.15)$$
where we have used eq.~(A.2). In fact, we just proved the identity
$$-C^{\pm~[ij}_{ac}C^{\pm ~kl]}_{bc}-2C^{\pm}_{a[ijk}\d_{l]b}
 -2C^{\pm}_{b[ijk}\d_{l]a}=\d_{ab}C^{\pm ijkl}~.$$
The explicit formulas for the gamma matrices will be used in the next section.

{} The matrices $\g^{ij}$ represent the 21 generators $J^{ij}$ of $Spin(7)$ in
its eight-dimensional spinor representation. One can extend the spinor 
representation of $Spin(7)$ to the left handed or right handed spinor 
representation of $SO(8)$ by adding the matrices $ \pm i\g^i$ \cite{gg,wn,dgt}.
By defining  $J^i=J^{i8}$, the commutation relations of $SO(8)$ can be written
as 
$$\eqalign{
\[ J^i,J^j\] =~&~ 2J^{ij}~,\cr
\[J^i,J^{mn}\]=~&~ 2\d^{im}J^n -  2\d^{in}J^m~,\cr
\[ J^{ij},J^{kl}\]=~&~ 2\d^{jk}J^{il} +  2\d^{il}J^{jk} -  2\d^{ik}J^{jl}
- 2\d^{jl}J^{ik}~.\cr}\eqno(2.16)$$
 The automorphism group $G_2$ of octonions is a 
14-dimensional subgroup of $SO(7)$. Under $G_2$, the adjoint representation of
$SO(7)$ decomposes as  $\ul{21}=\ul{14}+\ul{7}$.  We shall denote the
 generators of $G_2$ as 
 $G^{ij}$. One can choose a basis for $G_2$ such that the generators $G^{ij}$ 
can be expressed in terms of the generators $J^{ij}$ of $SO(7)$ in a simple 
form \cite{wn,gn}:
$$ G^{ij} = \ha J^{ij} + \fracm{1}{8}C^{ij}_{~~kl} J^{kl}~.\eqno(2.17)$$
Eq.~(2.17) implies the linear relations
$$ C_{ijk}G^{jk}=0~,\eqno(2.18)$$
and these are just the seven constraints that enforce the generators $G^{ij}$
to span the 14-dimensional vector space \cite{gg}. Note also the related 
identities
$$C_{ijkl}G^{kl}=2G_{ij}~,\quad {\rm and} \quad 
C^{[ij}_p G^{k]p} =0~. \eqno(2.19)$$  

The remaining seven generators  of  $SO(7)$  can be chosen as
$$A^i=\ha C^{ijk} J^{jk}~,\eqno(2.20)$$
They are associated with the seven-dimensional coset space $SO(7)/G_2$. 
Therefore, we arrive at the decomposition \cite{gn}
$$J^{ij}=\fracm{4}{3}G^{ij} + \fracm{1}{3}C^{ijk}A^k~.\eqno(2.21)$$

The $G_2$ generators $G^{ij}$ satisfy the commutation relations \cite{gn}:
$$
\[ G^{ij},G^{kl}\]= 2\d^{l[i}G^{j]k} -  2\d^{k[i}G^{j]l} + \ha\left(
C^{klm[i}G^{j]m} - C^{ijm[k}G^{l]m}\right)~,\eqno(2.22)$$
Furthermore, we have 
$$ \[ A^i, A^j\] = -8G^{ij} + 2 C^{ijk}A^k~,\eqno(2.23)$$
thus reflecting the fact that the coset space $SO(7)/G_2$ is {\it not} a 
symmetric space. The symmetric space $SO(8)/SO(7)$  can be identified with the
 round seven-sphere  $S^7$ . The space $SO(7)/G_2$  can be considered as the   
seven-sphere with torsion, and we shall denote it in what follows as 
$S^7_{\rm T}$. 

The $SO(8)$ generators can similarly be decomposed with respect to $G_2$,
$$ \ul{28} = \ul{14} + \ul{7} + \ul{7}~,\eqno(2.24)$$
with the generators $J^i$ and $A^i$ introduced above transforming in the 
seven-dimensional representation of $G_2$ . In a $G_2$ basis, the commutation 
relations of $SO(8)$ take the form
$$\eqalign{
\[ J^i,J^j\]= ~&~ 2J^{ij} = \fracm{8}{3}G^{ij} + \fracm{2}{3}C^{ijk}A^k~,\cr
\[A^i,J^j\] = ~&~ -2C^{ijk}J^k~,\cr
\[J^i,G^{kl}\] = ~&~ \d^{ik}J^l -\d^{il}J^k +\ha C^{iklp}J^p~,\cr
\[A^i,G^{kl}\] = ~&~ \d^{ik}A^l -\d^{il}A^k +\ha C^{iklp}A^p~,\cr }
 \eqno(2.25)$$
in addition to the commutation relations (2.16), (2.22) and (2.23).

The three eight-dimensional representations of $SO(8)$ are in triality and the 
subgroup of $SO(8)$ invariant under the triality mapping is $G_2$ ~\cite{gg}.
This is evident from the commutation relations of $SO(8)$ in the $G_2$ basis
above. Note also the following additional identities:
$$\eqalign{
C_{~~kl}^{ij}J^{kl} = ~&~ \fracm{8}{3}G^{ij} - \fracm{4}{3}C^{ijp}A^p~,\cr
J_{ij}J^{ij}= ~&~ \fracm{16}{9}G_{ij}G^{ij} + \fracm{2}{3}A_iA^i~,\cr
C_{ijkl}J^{ij}J^{kl} =~&~ \fracm{32}{9}G_{ij}G^{ij} - \fracm{8}{3}A_iA^i=
2J_{ij}J^{ij} - 4A_iA^i~.\cr}\eqno(2.26)$$

Another embedding of $G_2$ into $SO(7)$ that will be also usefull in the next 
sections was given in ref.~\cite{gg} and later used in ref.~\cite{bo}. In this
embedding the fourteen generators $M^A$ of $G_2$ $(A=1,2,..,14)$ are given as
follows:   
$$\eqalign{
M^1=\fracmm{1}{\sqrt{2}}\left(T^{41}+T^{36}\right)~,\quad & \quad
M^2=\fracmm{1}{\sqrt{6}}\left(T^{41}-T^{36}+2T^{25}\right)~,\cr
M^3=\fracmm{1}{\sqrt{2}}\left(T^{31}-T^{46}\right)~,\quad & \quad
M^4=\fracmm{1}{\sqrt{6}}\left(T^{31}+T^{46}-2T^{57}\right)~,\cr
M^5=\fracmm{1}{\sqrt{2}}\left(T^{21}-T^{76}\right)~,\quad & \quad
M^6=\fracmm{1}{\sqrt{6}}\left(T^{21}+T^{56}-2T^{45}\right)~,\cr
M^7=\fracmm{1}{\sqrt{2}}\left(T^{71}+T^{26}\right)~,\quad & \quad
M^8=\fracmm{1}{\sqrt{6}}\left(T^{71}-T^{26}-2T^{35}\right)~,\cr
M^9=\fracmm{1}{\sqrt{2}}\left(T^{24}-T^{73}\right)~,\quad & \quad
M^{10}=\fracmm{1}{\sqrt{6}}\left(T^{24}+T^{73}+2T^{15}\right)~,\cr
M^{11}=\fracmm{1}{\sqrt{2}}\left(T^{74}+T^{23}\right)~,\quad & \quad
M^{12}=\fracmm{1}{\sqrt{6}}\left(T^{74}-T^{23}-2T^{65}\right)~,\cr
M^{13}=\fracmm{1}{\sqrt{2}}\left(T^{43}-T^{16}\right)~,\quad & \quad
M^{14}=\fracmm{1}{\sqrt{6}}\left(T^{43}+T^{16}+2T^{27}\right)~,\cr}
\eqno(2.27)$$
where $T^{ij}$ are $SO(7)$ generators. For writing down the non-linear $N=7$ 
superconformal algebra (sect.~3), it is convenient to take the $SO(7)$
generators here in the vector representation \cite{bo}, 
$$(T^{ij})_{kl}= -i \left(\d^i_k\d^j_l - \d^i_l\d^j_k \right)~.\eqno(2.28)$$

Finally, we give a few branching rules for the $Spin(7)$ tensor products,
namely
$$\ul{7}\times\ul{7} = \ul{1}_{\rm \,s} + \ul{21}_{\rm \,a} 
+ \ul{27}_{\rm \,s}~, \quad
\ul{8}\times\ul{7} = \ul{8} + \ul{48}~,\quad \ul{8}\times \ul{8} = 
\ul{1}_{\rm \,s} + \ul{7}_{\rm \,a} + \ul{21}_{\rm \,a} + 
\ul{35}_{\rm \,s}~,\eqno(2.29)$$
where the $\ul{8}$ stands for the 8-dimensional spinor representation. 
As far as $G_2$ is concerned, the only decomposition to be relevant for us is 
given by
$$\ul{7}\times \ul{7} = \ul{1} + \ul{7} + \ul{14} + \ul{27}~.\eqno(2.30)$$
\vglue.2in

In Table I, we list some basic facts about $G_2, SO(7)$ and $SO(8)$. 
\begin{center}
{\sf Table I}. The groups $G_2$, $SO(7)$ and $SO(8)$
\vglue.1in
\noindent\begin{tabular}{cccc} \hline
$G$ & dimension & rank & dual~Coxeter~number \\
 \hline
$G_2$ &    14         &      2     &            4 \\
$SO(7)$ &  21         &      3     &            5\\
$SO(8)$ &  28         &      4     &            6\\
\hline
\end{tabular}
\end{center}

\vglue.3in

\section{Exceptional non-linear superconformal algebras}

In this section, we present the defining OPEs for the $N=8$ and $N=7$
non-linear superconformal algebras following ref.~\cite{bo}. These algebras 
can be obtained either via a Drinfeld-Sokolov-type reduction from affine 
versions of the exceptional Lie superalgebras $F(4)$ and $G(3)$, respectively
\cite{imp}, or by purely algebraic methods~\cite{fl,bo}. Both algebras have 
generators of conformal dimension $2$, $3/2$ and $1$ only. The $N=8$ algebra 
contains eight supercurrents $S^M$ of conformal dimension $3/2$, and $21$ 
symmetry currents of $SO(7)$ under which the supercurrents transform in the 
{\it spinor} representation. The $N=7$ algebra has $7$ supercurrents, and $14$
 symmetry currents of $G_2$. Both algebras contain a single generator of 
conformal dimension $2$, and they are {\it completely} fixed by their field 
content and associativity (the Jacobi `identities'). Because of  their 
non-linearity, the `vacuum-preserving' algebra, generated by the modes 
$L_0$, $L_{\pm 1}$, $S^M_{\pm 1/2}$ and $T^A_0$, is {\it not\/} finite. The 
OPEs to be given below  are equivalent to the (anti)commutation relations  of
ref.~\cite{bo}.
\vglue.2in

\subsection{Exceptional ~$N=8$~  superconformal algebra}

The bosonic part of the $N=8$ algebra is a semi-direct product of the affine
algebra $\Hat{so(7)}_k$ of level $k$ and the Virasoro algebra. The 
corresponding OPEs are given by 
$$\eqalign{
T(z)\;T(w)~\sim~&~\fracmm{c/2}{(z-w)^4} + \fracmm{2T(w)}{(z-w)^2} +
\fracmm{\pa T(w)}{z-w}~,\cr
T(z)T^{mn}(w)~\sim~&~\fracmm{T^{mn}(w)}{(z-w)^2} + \fracmm{\pa T^{mn}(w)}{z-w}
~,\cr}\eqno(3.1)$$
and
$$\eqalign{
T^{mn}(z)T^{pq}(w)~\sim~&~ \fracmm{-i}{z-w}\left\{ \d^{np}T^{mq}(w) +
 \d^{mq}T^{np}(w)  - \d^{mp}T^{nq}(w) -  \d^{nq}T^{mp}(w) \right\} \cr
{} ~&~ +\fracmm{k}{(z-w)^2}\left\{ \d^{mp}\d^{nq} -  \d^{mq}\d^{np}\right\}~,
\cr}\eqno(3.2)$$
where the adjoint of $SO(7)$ is labeled by a pair of antisymmetric indices, 
$m,n,\ldots=1,2,\ldots,7$. Compared to the previous section, we have normalised
the affine spin-1 currents differently, $J^{mn}=2iT^{mn}_0$, where $T^{mn}_0$
 is the zero-mode of $T^{mn}(z)$.~\footnote{The same normalisation convention
was adopted in ref.~\cite{bo}.}

Since the $N=8$ supercurrents $S^M(z)$ transform in the spinor representation 
of $SO(7)$ and have spin 3/2, we have 
$$\eqalign{
T(z)\;S^M(w)~\sim~&~\fracmm{\frac{3}{2}S^M(w)}{(z-w)^2}+\fracmm{\pa S^M(w)}{z-
w}~,\cr
T^{mn}(z)S^M(w)~\sim~&~\fracmm{-i}{2} \fracmm{\g^{mn}_{MN}}{(z-w)} S^N(w)~,\cr}
\eqno(3.3)$$

The only non-trivial OPE's are the ones corresponding to the products of
 $N=8$ supersymmetry generators which read as follows:
$$ S^M(z)S^N(w)~\sim~\fracmm{8k(k+2)}{3(k+4)}\fracmm{\d^{MN}}{(z-w)^3} +
\fracmm{2T(w)}{z-w}\d^{MN}
- \fracmm{\d^{MN}}{3(k+4)}\fracmm{:T^{mn}T^{mn}:(w)}{z-w} \eqno(3.4)$$
$$+ \fracmm{2i(k+2)}{3(k+4)}\g^{MN}_{mn}\left[ \fracmm{T^{mn}(w)}{(z-w)^2} +
 \fracmm{\pa T^{mn}(w)}{2(z-w)}\right] - \fracmm{1}{12(k+4)}\g^{MN}_{mnpq}
\fracmm{:T^{mn}T^{pq}:(w)}{z-w}~,$$
where $M,N = 1,\ldots,8$, and $i,j,\ldots=1,\ldots,7$, as in the previous
section. The above eq.~(3.4) becomes more transparent and suitable for 
calculations after substituting the gamma matrices in terms of the octonionic 
structure constants (see sect.~2 and Appendix A). The relevant formulas  are 
collected in Appendix B. 

We have verified that all the Jacobi `identities' are satisfied provided that 
the central charge $c$ of the $N=8$ algebra is determined by the level $k$ as
follows:
$$ c= c_8\equiv 4k + \fracmm{6k}{k+4} \equiv \fracmm{2k(2k+11)}{k+4}~,
\eqno(3.5)$$
in agreement with refs.~\cite{fl,bo}. The identities (2.8) and (2.9) were 
crucial in checking the Jacobi `identities' for the N=8 algebra. Compared to
Bowcock \cite{bo}, our supersymmetry generator $S^M$ above differs from his
, $S^M_{\rm B}$, by an overall scale factor, namely,
$S^M=\sqrt{\fracm{2k+4}{2k+11}}S^M_{\rm B}$.  It should be stressed that the 
very existence of such non-linear $N=8$ superconformal algebra is highly 
non-trivial, because several consistency requirements still have to be 
satisfied in the process of checking the Jacobi `identities' when all free 
parameters are already fixed \cite{bo}.  
\vglue.2in

\subsection{Exceptional ~$N=7$~  superconformal algebra}

The exceptional $N=7$~ non-linear superconformal algebra is similar to the 
$N=8$~ algebra, with the gauge group $G_2$ instead of $SO(7)$, and seven 
supercurrents. We shall denote the generators of $G_2$ as $G^A$ and not as
$M^A$ as we did in the previous section. The symbol $M^A$ will be used
for  the seven-dimensional representation matrices of $G_2$, $A=1,2,\ldots,14$,
 as given in  eq.~(2.27), providing the explicit embedding of $G_2$ into 
$SO(7)$ \cite{gg}. The matrices $M^A$ satisfy the properties \cite{bo}
$$\eqalign{
\tr\left(M^AM^B\right)~=~&~2\d^{AB}~,\cr
M^A_{ij}\;M^A_{kl}~=~&~ \fracmm{2}{3}\left(\d_{il}\d_{jk}-\d_{ik}\d_{jl}\right)
-\fracmm{1}{3}C_{ijkl}~.\cr}\eqno(3.6)$$

The bosonic OPEs of the  $N{=}7$~ algebra are
$$\eqalign{
T(z)\;T(w)~\sim~&~\fracmm{c/2}{(z-w)^4} +\fracmm{2T(w)}{(z-w)^2} +\fracmm{\pa
T(w)}{z-w}~,\cr
T(z)G^{ij}(w)~\sim~&~\fracmm{G^{ij}(w)}{(z-w)^2} 
+\fracmm{\pa G^{ij}(w)}{z-w}~.\cr}\eqno(3.7)$$
The $G_2$ currents satisfy~\cite{gn} 
$$\eqalign{
G^{ij}(z)G^{kl}(w)~\sim~&~\fracmm{1}{z-w}\left\{ 2\d^{l[i}G^{j]k}(w) -  
2\d^{k[i}G^{j]l}(w)\right. \cr
{}~&~ \left. + \fracm{1}{2}C^{klm[i}G^{j]m}(w) - \fracm{1}{2}
C^{ijm[k}G^{l]m}(w)\right\} \cr
{}~&~ -\fracmm{k}{(z-w)^2}\left\{ 
\frac{3}{2}\left(\d^{ik}\d^{jl}-\d^{il}\d^{jk}\right) +\frac{3}{4}C^{ijkl}
\right\}~.\cr}\eqno(3.8)$$
The generators $G^{ij}$ of $G_2$ are related to the generators $G^A$ given
in the previous section as
$$ G_{ij} = \fracmm{3}{4i} M_{ij}^A G^A~.\eqno(3.9) $$

The seven supercurrents $S^i(z)$ transform in $\ul{7}$ of $G_2$, and 
satisfy the OPEs
$$\eqalign{
T(z)\;S^i(w)~\sim~&~\fracmm{\frac{3}{2}S^i(w)}{(z-w)^2} 
+\fracmm{\pa S^i(w)}{z-w}~,\cr
 G^A(z)S^i(w)~\sim~&~\fracmm{1}{z-w}M^A_{ij}\,S^j(w)~.\cr}\eqno(3.10)$$

The most important OPE's are again the ones defining the  $N=7$ supersymmetry
algebra, and they read as
$$\eqalign{
S^i(z)S^j(w)~\sim~&~\fracmm{k(3k+5)}{k+3}\fracmm{\d^{ij}}{(z-w)^3}
+\fracmm{3k+5}{k+3}M^A_{ij}\left[\fracmm{G^A(w)}{(z-w)^2}+\ha \fracmm{\pa
G^A(w)}{z-w}\right]\cr
{}~&~+\fracmm{\d^{ij}}{z-w}\left[2T(w) - \fracmm{1}{k+3}:G^AG^A:(w)\right]\cr
{}~&~+\fracmm{3}{4(k+3)}\left[ M^AM^B +M^BM^A\right]^{ij}\fracmm{:G^AG^B:(w)}{
z-w}~.\cr}\eqno(3.11)$$

We have verified that all the Jacobi `identities' are indeed satisfied 
provided that the central charge is given by \cite{fl,bo}
$$ c = c_7 \equiv \fracmm{9}{2}k + \fracmm{2k}{k+3} \equiv 
\fracmm{k(9k+31)}{2(k+3)}~.\eqno(3.12)$$
Again, it is fully consistent, in particular, with the results of Bowcock 
\cite{bo}, after taking into account the rescaling 
$S^i=\sqrt{\fracm{9k+15}{9k+31}}S^i_{\rm B}$.

There is a confusion in the literature concerning the relationship between the
two exceptional superconformal algebras. The $N=7$ non-linear algebra is 
{\it not} a subalgebra of the $N{=}8$ non-linear algebra. This can be most 
easily seen in the limit $c\to\infty$, where both algebras linearise. 
The $N=8$ algebra contains the finite Lie superalgebra $F(4)$ in that limit, 
which is its vacuum-preserving subalgebra. On the other hand, the 
corresponding subalgebra of the $N=7$ algebra is $G(3)$. If the  $N=7$ algebra
were a subalgebra of the $N=8$ one, then $G(3)$ would have to be a subalgebra 
of $F(4)$. However, it is known that this is not the case. It follows from the
fact that the smallest non-trivial representations of both Lie superalgebras 
are their adjoint representations. If $G(3)$ were a subalgebra of $F(4)$, then
this would  imply  that there be an 9-dimensional 
$({\rm dim~ad}\,F(4)-{\rm dim~ad}\,G(3)=9)$ non-trivial 
 representation of $G(3)$, which does not exist \cite{kac}.
\vglue.2in

\section{Exceptional coset constructions}

In this section, we shall investigate the possibility of realizing the 
exceptional superconformal algebras over certain special coset spaces $G/H$.
We adopt here the following conventions:~\footnote{The notation adopted here, 
in this section, for a {\it general} group $G$ and its subgroup $H$ slightly 
overlaps with our conventions in the previous sections and in what follows for
the particular cosets. This should not lead to a confusion since we 
discuss the general and particular cosets separately in our paper.} we
use the early Latin capital letters for $G$ indices, and the early lower-case 
Latin letters for $G/H$ indices, $A,B,\ldots=1,\ldots,{\rm dim}\,G$, and  
$a,b,\ldots={\rm dim}\,H + 1,\ldots,{\rm dim}\,G$.

Let $\tilde{k}$ be (integer) level of affine algebra $\Hat{G}$ realised in 
terms of (bosonic) currents $\hat{J}^A(z)$. The latter can be thought of as
originating from the bosonic WZNW model on the group $G$ \cite{wznw,jb,book},
and they satisfy the OPE 
$$\hat{J}^A(z)\hat{J}^B(w) = \fracmm{\tilde{k}/2}{(z-w)^2}\d^{AB} 
+ \fracmm{if^{ABD}\hat{J}^D(w)}{z-w} + :\hat{J}^A\hat{J}^B:(w)
 + \ldots~,\eqno(4.1)$$
where $if^{ABD}$ are the structure constants of $G$. Let's also introduce free
fermions in the adjoint of $G$, with the OPE
$$\j^A(z)\j^B(w) = \fracmm{1/2}{z-w}\d^{AB} + :\j^A\j^B:(w) + 
(z-w):\pa\j^A\j^B:(w) + \ldots~.\eqno(4.2)$$
These free fermions can be thought of as coming from the (1,0) {\it 
supersymmetric} (heterotic) WZNW model on the group $G$ \cite{vkpr,book}.

The basic idea of coset construction  is to construct generators of a given
superconformal algebra in terms of the basic fields $\hat{J}^a(z)$ and 
$\j^a(z)$ associated with a coset $G/H$ \cite{gko,jb,book}. Eqs.~(4.1) and 
(4.2) obviously imply
$$\hat{J}^a(z)\hat{J}^b(w) = \fracmm{\tilde{k}/2}{(z-w)^2}\d^{ab}
+ \fracmm{if^{abD}\hat{J}^D(w)}{z-w} \, + :\hat{J}^a\hat{J}^b:(w)
 + \ldots~,\eqno(4.3)$$
and
$$\j^a(z)\j^b(w) = \fracmm{1/2}{z-w}\d^{ab} + :\j^a\j^b:(w) +
(z-w):\pa\j^a\j^b:(w) + \ldots~.\eqno(4.4)$$

The most general Ansatz for the supercurrents of any extended superconformal 
algebra over an arbitrary coset is given by
$$ S^M(z) = 2\a(\tilde{k}) \left\{ h^M_{ab}\j^a(z)\hat{J}^b(z) 
+  \g(\tilde{k}) \x^M_a\pa\j^a(z)
- \fracmm{2i}{3}\b(\tilde{k})\G^M_{abc}:\j^a\j^b\j^c:(z)
 \right\}~,\eqno(4.5)$$
where $\a(\tilde{k})$, $\g(\tilde{k})$ and $\b(\tilde{k})$ are some functions 
of the level $\tilde{k}$, while $h^M_{ab}$, $\x^M_a$ and $\G^M_{abc}$ are some
tensors, the latter being totally antisymmetric with respect to its 
subscripts.~\footnote{No symmetry properties are \`a priori assumed for 
$h^M_{ab}$ and $\x^M_a$.} 

The Ansatz (4.5) is dictated by dimensional reasons. The tensors $h^M_{ab}$, 
$\x^M_a$ and $\G^M_{abc}$ have to be consistent with the transformation 
properties of the conformal fields in the Ansatz (4.5). For example, the 
`background charge' terms proportional to $\pa\j^a$ can only contribute when 
there exists a mixed tensor $\x^M_a$ which is invariant under the $SO(7)$ 
transformations in the N=8 case or under the $G_2$ transformations in the N=7 
case. That is only possible if this tensor is proportional to the delta-symbol, 
which implies, in particular when $\g\neq 0$, that some free fermions $\j^a$ 
should transform in the {\it same} $SO(7)$ or $G_2$ representation as the N=8 
or N=7 supercurrents, respectively. 

The tensors  $h^M_{ab}$ can be geometrically interpreted as the generalised 
complex structures on the coset in question, whereas $\G^M_{abc}$ as the 
generalized torsion coefficients. The tensor $\x^M_a$ represents the background
charges (see subsect.~4.2 for a non-trivial example).

The Ansatz (4.5) is supposed to be completely fixed by the superconformal 
algebra. This is known to be the case for the superconformal algebras with 
$N\leq 4$ \cite{gko,ks,vp,st,mg93,gk}, and it is expected to be the case in 
general. In fact, the resulting constraints usually lead to an overdetermined 
system of equations, so that it is highly non-trivial whether the equations 
are really consistent and lead to a solution when the number of supersymmetries
 is larger than four.  As we shall show below, there exist 
very few consistent solutions to these constraints in the case of the 
exceptional $N=7$ and $N=8$ extended superconformal algebras.

For the unitary representations to be constructed via the coset space method, 
the coefficients on the r.h.s. of eq.~(4.5) for the supercurrents have to be 
real, if the fermions are normalised as above and the currents $\hat{J}^a$ are
hermitian.

It is straightforward to calculate the OPE that the supercurrents (4.5)
 satisfy.
We find
$$\eqalign{
(4\a^2)^{-1}S^M(z)S^N(w)\sim &~\fracmm{1}{(z-w)^3}\left[ \fracm{\tilde{k}}{4}
h^M_{ab}h^N_{ab} + \fracm{1}{3}\b^2 \G^M_{abc}\G^N_{abc} - \g^2\x^M_a\x^N_a
\right] \cr
{}~&~ + \fracmm{1}{(z-w)^2} \left[ \fracm{i}{2}h^M_{ab}h^N_{ad}f^{bdD}
\hat{J}^D + \fracm{\tilde{k}}{2}h^M_{ab}h^N_{cb}:\j^a\j^c: \right. \cr
{}~&~ + 2\b^2\G^M_{abc}\G^N_{abd} :\j^c\j^d: 
 + \g\ha(h^M_{cb}\x^N_c - h^N_{cb}\x^M_c)\hat{J}^b  \cr
{}~&~ \left. -2i\b\g(\G^M_{abc}\x^N_a-\G^N_{abc}\x^M_a):\j^b\j^c: \right](w)\cr
{}~&~  + \fracmm{1}{z-w} \left[
 +\fracm{1}{2} h^M_{ab}h^N_{ad}(:\hat{J}^b\hat{J}^d:+ \fracm{1}{2} if^{bdD}
\pa\hat{J}^D) + \ha \g h^M_{ab}\x^N_a\pa\hat{J}^b  \right. \cr
{}~&~  + 2\b^2\G^M_{abf}\G^N_{abg}:\pa\j^f\j^g: 
-i\b\left(\G^M_{abc}h^N_{cf} + \G^N_{abc}h^M_{cf}\right)
\hat{J}^f:\j^a\j^b:\cr
{}~&~ -4i\b\g\G^M_{abc}\x^N_a:\j^b\pa\j^c: 
+ ih^M_{ab}h^N_{cd}f^{bdD}\hat{J}^D:\j^a\j^c:  \cr
{}~&~\left. + \fracm{\tilde{k}}{2}h^M_{ab}h^N_{cb}:\pa\j^a\j^c:
 -2\b^2\G^M_{abc}\G^N_{afg}:\j^b\j^c\j^f\j^g:\right](w)~.\cr}
\eqno(4.6)$$

As far as the $N=8$ algebra is concerned, comparing the residues at the 
$(z-w)^{-3}$, $(z-w)^{-2}$ and $(z-w)^{-1}$ poles in eqs.~(3.4) and (4.6), 
respectively, yields  
$$\fracmm{8k(k+2)}{3(k+4)}\d^{MN} = \a^2 \left[ \tilde{k}h^M_{ab}h^N_{ab} 
+ \fracm{4}{3}\b^2 \G^M_{abc}\G^N_{abc} - 4\g^2\x^M_a\x^N_a 
\right]~,\eqno(4.7a)$$
$$\eqalign{
\fracmm{2i(k+2)}{3(k+4)}\g^{MN}_{mn}T^{mn}(z) =~&~ \a^2 \left[ 
 2ih^M_{ab}h^N_{ad}f^{bdD}\hat{J}^D + 4\g h^{[M}_{ab}\x^{N]}_a\hat{J}^b
 - 16i \b \g \G^{[M}_{abc}\x^{N]}_a:\j^b\j^c: \right. \cr
{}~&~ + \left. 2\tilde{k}h^M_{ab}h^N_{cb}:\j^a\j^c: +  
 8\b^2\G^M_{abc}\G^N_{abd} :\j^c\j^d:\right](z)~,\cr}\eqno(4.7b)$$
and
$$
\left[ 2T(z) - \fracmm{1}{3(k+4)}:T^{mn}T^{mn}:(z)\right] \d^{MN} + 
\fracmm{i(k+2)}{3(k+4)}\g^{MN}_{mn}\pa T^{mn}(z) $$
$$ - \fracmm{1}{12(k+4)}\g^{MN}_{mnpq}:T^{mn}T^{pq}:(z) = 
 \a^2 \left[ 2 h^M_{ab}h^N_{ad} \left( :\hat{J}^b\hat{J}^d: 
+ \fracm{1}{2} if^{bdD}\pa\hat{J}^D\right)(z) \right. $$  
$$  + 4ih^M_{ab}h^N_{cd}f^{bdD}\hat{J}^D:\j^a\j^c:(z) 
 -4i\b\left(\G^M_{abc}h^N_{cf} + \G^N_{abc}h^M_{cf}\right)
\hat{J}^f:\j^a\j^b:(z)$$
$$ + 2\tilde{k} h^M_{ab}h^N_{cb}:\pa\j^a\j^c:(z)
 + 8\b^2\G^M_{abf}\G^N_{abg}:\pa\j^f\j^g:(z) +2\g h^M_{ab}\x^N_a
\pa\hat{J}^b(z)$$
$$\left. - 16i\b \g\G^M_{abc}\x^N_a:\j^b\pa\j^c:(z)
-8\b^2\G^M_{abc}\G^N_{afg}:\j^b\j^c\j^f\j^g:(z)\right]~.
\eqno(4.7c)$$
where  $M,N=1,2,\ldots,8$.

Quite similar equations appear in the case of the $N{=}7$ algebra. We find
$$\fracmm{k(3k+5)}{k+3}\d^{mn} = \a^2 \left[ \tilde{k}h^m_{ab}h^n_{ab}
+ \fracm{4}{3}\b^2 \G^m_{abc}\G^n_{abc} - 4\g^2\x^m_a\x^n_a \right]~,
\eqno(4.8a)$$
$$\eqalign{
\fracmm{3k+5}{k+3}\left(M^A\right)^{mn}G^A(z) =~&~ \a^2 \left[
 2ih^m_{ab}h^n_{ad}f^{bdD}\hat{J}^D(z) + 4\g h^{[m}_{ab}\x^{n]}_a\hat{J}^b
- 16i \b \g \G^{[m}_{abc}\x^{n]}_a:\j^b\j^c: \right. \cr
{}~&~ + \left. 2\tilde{k}h^m_{ab}h^n_{cb}:\j^a\j^c:(z) +
 8\b^2\G^m_{abc}\G^n_{abd} :\j^c\j^d:(z)\right]~,\cr}\eqno(4.8b)$$
and
$$
\left[ 2T(z) - \fracmm{1}{k+3}:G^AG^A:(z)\right] \d^{mn} +
\fracmm{3k+5}{2(k+3)}\left(M^A\right)^{mn} \pa G^A(z) $$
$$ + \fracmm{1}{4(k+3)} \left(M^AM^B + M^BM^A\right)^{mn}:G^AG^B:(z) =
 \a^2 \left[ 2 h^m_{ab}h^n_{ad}\left(:\hat{J}^b\hat{J}^d:+
\fracm{1}{2} if^{bdD}\pa\hat{J}^D
\right)(z) \right.$$
$$    + 2\tilde{k} h^m_{ab}h^n_{cb}:\pa\j^a\j^c:(z) 
 + 4ih^m_{ab}h^n_{cd}f^{bdD}\hat{J}^D:\j^a\j^c:(z) 
+2\g h^m_{ab}\x^n_a\pa\hat{J}^b(z) $$
$$ -4i\b\left(\G^m_{abc}h^n_{cf} + \G^n_{abc}h^m_{cf}\right)
\hat{J}^f:\j^a\j^b:(z) + 8\b^2\G^m_{abf}\G^n_{abg}:\pa\j^f\j^g:(z)$$
$$\left.  -16i\b \g\G^m_{abc}\x^n_a:\j^b\pa\j^c:(z)
-8\b^2\G^m_{abc}\G^n_{afg}:\j^b\j^c\j^f\j^g:(z)\right],~
\eqno(4.8c)$$
where $m,n=1,2,\ldots,7$, and $A=1,2,\ldots,14$.

For  both $N=7$ and $N=8$ algebras, eq.~(a) determines, in particular, the 
level $k$ of the algebra, eq.~(b) determines the affine currents of the 
algebra, while eq.~(c) determines the stress tensor. Furthermore,  for 
each equation, there are the complicated non-linear consistency conditions on
the unknown constant tensors $h$, $\x$ and $\G$, and the unknown coefficients 
$\a$, $\b$ and $\g$. For example, as far as the $(M,N)$-{\it symmetric} 
simple-pole contributions on the r.h.s. of eq.~(4.7c) for the $N=8$ algebra 
are concerned, the coset current ($\hat{J}^a$)-dependent terms among them are 
given by
$$ h^{(M}_{ab}h^{N)}_{ad}:\hat{J}^b\hat{J}^d:-4i\b\G^{(M}_{abc}h^{N)}_{cf}
\hat{J}^f :\j^a\j^b: + \g h^{(M}_{ab}\x^{N)}_a\pa\hat{J}^b~.\eqno(4.9)$$ 
They can only contribute to the trace ($\d^{MN}$-dependent) terms, according 
to eq.~(4.7c). Moreover, the term quadratic in the coset currents  has to be 
diagonal (i.e. of Sugawara form), since it is going to contribute to the 
stress tensor $T$. Therefore, we conclude  that   
$$ h^{(M}_{ab}h^{N)}_{ad}\sim \d^{MN}\d_{bd}~,\qquad 
\G^{(M}_{abc}h^{N)}_{cf}\sim \d^{MN}~, \quad {\rm and } \quad
h^{M}_{ab}\x^{N}_a\sim \d^{MN}~, \eqno(4.10)$$
where, in the last equation, we have also taken into account the restrictions
coming from the antisymmetric terms in eq.~(4.7c) too. The conditions (4.10)
are highly restrictive since, in addition, all the tensors $h$, $\G$ and $\x$ 
are to be consistent with the transformation properties of the both sides of 
eq.~(4.5).

The first equation (4.10) implies that $h^m_{ab}$ must be the 
seven-dimensional $8\times 8$ gamma matrices and $h^8\sim {\bf 1}_8$ 
if we assume that the indices $a,b,\ldots$ take values in an irreducible 
representation of $SO(7)$ (by Schur's lemma).~\footnote{The naive solution 
-- the eight-dimensional $16\times 16$ gamma matrices --- is ruled out 
because it leads to  $SO(8)$ gauge invariance instead of  $SO(7)$ 
required.} The second equation (4.10) then becomes equivalent to the relation
$$ \G^m_{abc}=ih^m_{cd}\G^8_{abd}~.\eqno(4.11)$$
It is not difficult to verify that eq.~(4.11) does not have a non-trivial 
solution for $\G^M_{abc}$ which whould be totally antisymmetric with respect 
to its subscript indices, as required. Hence, we have to conclude that either 
all $\hat{J}^m$ {\it or\/} all $\G^M$ have to vanish. Similar conclusions 
follow for the N=7 case too (see subsect.~4.2).

Thus, the coset we are looking for, if any, should be $(7+1)$-dimensional, 
the seven-dimensional space being represented by a seven-sphere (the only 
parallelizable coset space in seven dimensions). Indeed, the very existence of
the exceptional $N=8$ and $N=7$ algebras crucially depends upon unique 
properties of gamma matrices in seven dimensions, which are related to 
octonions. One should therefore have expected that the coset spaces in 
question are to be the ones given by various symmetry groups of octonions. The
naive candidates for such cosets would be $SO(8)/SO(7)$ for the $N=8$ algebra,
which corresponds to the round seven-sphere $S^7$, and $SO(7)/G_2$ for the 
$N=7$ algebra, which correponds to the $S^7_T$ with torsion. However, it is 
not difficult to convince oneself that these naive coset spaces can {\it not} 
be the right ones. For the $N=8$ algebra, we need supercurrents that transform 
in the spinor representation of $SO(7)$, which can not be obtained from the 
currents and fermions on $S^7$ that transform in the vector representation of 
$SO(7)$, according to eq.~(2.29). For the $N=7$ algebra, the naive guess fails
due to the fact that $SO(7)/G_2$ is not a symmetric space, which leads to some 
unwanted terms in the OPEs. Remarkably enough, a simple extension of the naive 
coset spaces by a $U(1)$ factor, i.e. adding a circle or `1-sphere' $S^1$,  
leads to the consistent solutions for the above constraints (4.7) and (4.8),
as we are going to demonstrate in the rest of the paper. Adding the $U(1)$
current $J(z)$ is equivalent to introducing a scalar field $\f(z)$ since
$J\sim \pa\f$ up to a normalisation constant ({\it cf.}\/ ref.~\cite{m1} 
where the free-field representations for the orthogonal series of the 
Bershadsky-Knizhnik non-linear $N$-extended superconformal algebras were 
constructed).
\vglue.2in

\subsection{A construction of the exceptional $N=8$ superconformal
algebra over the coset space  ~$SO(8)\times U(1)/SO(7)$}

Our starting point is the affine algebra $\Hat{so(8)}_{\hat{k}} \oplus 
\Hat{u(1)}$, defined by the OPEs 
$$\eqalign{
\hat{J}^{ab}(z)\hat{J}^{cd}(w)~\sim~&~ 
\fracmm{2}{z-w}\left\{ \d^{bc}\hat{J}^{ad}(w)
 + \d^{ad}J^{bc}(w)  - \d^{ac}\hat{J}^{bd}(w) -  \d^{bd}\hat{J}^{ac}(w) 
\right\} \cr
{} ~&~ -\fracmm{4\hat{k}}{(z-w)^2}\left\{ \d^{ac}\d^{bd}-\d^{ad}\d^{bc}
\right\}~,\cr}\eqno(4.12)$$
and
$$\hat{J}^8(z)\hat{J}^8(w) ~\sim~\fracmm{\hat{k}_1/2}{(z-w)^2}~,\eqno(4.13)$$
where $a,b,\ldots=1,2,\ldots,8$, and $\hat{k}_1$ is a normalisation parameter
of the $U(1)$ current $\hat{J}^8$. The latter can be represented in terms of a 
scalar field, $\hat{J}^8\equiv i\sqrt{\hat{k}_1/2}\,\pa\f$.

Because of eq.~(4.12), the currents $\hat{J}^m=\pm i\hat{J}^{m8}$, 
$m=1,\ldots,7$, satisfy the OPE
$$\hat{J}^m(z)\hat{J}^n(w) = \fracmm{4\hat{k}}{(z-w)^2}\d^{mn} +
\fracmm{2\hat{J}^{mn}(w)}{z-w}\, +:\hat{J}^m\hat{J}^n:(w)+\ldots~.\eqno(4.14)$$

Associated with the $U(1)$ factor is an additional free fermionic field 
$\j^8(z)$, with the OPE
$$ \j^8(z)\j^8(w) = \fracmm{1/2}{z-w} + (z-w):\pa\j^8\j^8:(w) +\ldots~.
\eqno(4.15)$$
This field $\j^8$ together with the fermions $\j^m$ form an 8-dimensional
column $\j^a$, with the OPE (4.4). The currents $\hat{J}^m$ transform in
${\underline{7}}$ of $SO(7)$, while  $\hat{J}^8$ is a singlet. The  $\j^a$ 
will transform in the ${\underline{8}}$ of a $Spin(7)$ algebra to be 
constructed from fermion bilinears. 

Being applied to the particular coset space $SO(8) \times U(1)/SO(7)$, our 
general Ansatz (4.5) for the  $N=8$ supercurrents can be simplified to
$$\eqalign{
S^m(z) =~&~ 2\a \left\{ \g^m_{an}\j^a(z)\hat{J}^n(z) +i\j^m(z)\hat{J}^8(z)
+\g\pa\j^m(z) - \fracmm{2i}{3}\b \G^m_{abc}:\j^a\j^b\j^c:(z)\right\}~,\cr
S^8(z) =~&~ 2\a \left\{ -i\j^n(z)\hat{J}^n(z) -i\j^8(z)\hat{J}^8(z)
+\g\pa\j^8(z)- \fracmm{2i}{3}\b \G^8_{abc}:\j^a\j^b\j^c:(z)\right\}~,\cr}
\eqno(4.16)$$
where the parameter $\g=\g(\hat{k})$ plays the role of a background charge.
To simplify the structure of our Ansatz (4.16) even further, we represent the 
generalised complex structures in terms of the `extended' gamma matrices to be
 defined as
$$ h^M_{ab} = \g^M_{ab}~,\quad M=1,2,\ldots,8~,\quad {\rm with} \quad
\g^8 \equiv -i{\bf 1}_8~,\eqno(4.17)$$
These matrices satisfy the identities
$$ \g^8_{ab}\g^8_{ab}=-8~,\quad  \g^m_{ab}\g^8_{ab}=0~,\quad
 \g^M_{ab}\g^N_{ab}=-8\d^{MN}~,\eqno(4.18a)$$
and
$$ \eqalign{
\g^m_{na}\j^a\hat{J}^m + \g^8_{na}\j^a\hat{J}^8 = \g^n_{ab}\j^a\hat{J}^b~,\cr
\g^m_{8a}\j^a\hat{J}^m + \g^8_{8a}\j^a\hat{J}^8 = \g^8_{ab}\j^a\hat{J}^b~,\cr}
\eqno(4.18b)$$
and allow us to rewrite eq.~(4.16) as follows:
$$ S^M(z) = 2\a(\hat{k}) \left\{ \g^M_{ab}\j^a(z)\hat{J}^b(z) + \g(\hat{k})
\pa\j^M(z)
- \fracmm{2i}{3}\b(\hat{k}) \G^M_{abc}:\j^a\j^b\j^c:(z)\right\}~.\eqno(4.19)$$

The only choice for the generalised torsion coefficients $\G^M_{abc}$ on a
seven-sphere is  
$$ \G^8_{mnp}=\bar{A}C_{mnp}~,\quad  \G^{m}_{npq}=\bar{B}C\ud{m}{npq}~,\quad
 \G^m_{np8}=\bar{C}C\ud{m}{np}~,\eqno(4.20)$$
where the coefficients $\bar{A}$, $\bar{B}$ and $\bar{C}$ are at our disposal.

We thus reduce the problem of a coset space realization for the 
N=8 algebra to finding a solution for the coefficients 
$(\a,\b,\g,\hat{k}_1,\bar{A},\bar{B},\bar{C})$ from a consistency of the 
OPE (4.6) in terms of our Ansatz supercurrents (4.19) with the OPEs of the 
$N=8$ algebra. 
In particular, the r.h.s. of eq.~(4.7a) must reproduce $\d^{MN}$ and
 determine the  level $k$. Eq.~(4.7b) can be used to determine the
$SO(7)$ affine currents $T^{mn}(z)$ of the $N=8$ algebra: taking $M=m$ and 
$N=n$, where $m,n=1,2,\ldots,7$, we obtain the expressions for $T^{mn}(z)$ 
and, hence, $C\ud{p}{mn}T^{mn}(z)$, after using eqs.~(2.14), (4.18) and 
(4.20), and the identities of Appendix A. On the other hand, taking $M=m$ and 
$N=8$ in the same eq.~(4.7b), we can directly calculate $C\ud{p}{mn}T^{mn}(z)$.
 Both results must agree, and this gives us a non-trivial consistency relation.
The simple-pole contributions of eq.~(4.7c) produce three equations: the trace
 part proportional to $\d^{MN}$ determines the stress tensor $T$ of the $N=8$
algebra, whereas the antisymmetric part and the traceless symmetric part 
proportional to $\g_{mn}^{MN}$ and $\g_{mnpq}^{MN}$, 
respectively, yield the consistency relations for the already determined 
operators $\pa T^{mn}$ and $:T^{mn}T^{pq}:\,$.

The most severe restrictions come out of the symmetric traceless part of the 
simple-pole terms. First of all, the coset current ($\hat{J}^M$-dependent)
 contributions have to cancel, since they are 
obviously not allowed to contribute to a bilinear in the $SO(7)$ currents. 
There are two different types of such unwanted terms on the r.h.s. of 
eq.~(4.7c). First, the $\g$-dependent contribution contains the term  
$\g\frac{1}{2}\g^M_{Nb}\pa\hat{J}^b$ 
which has to be proportional to $\d^M_N$. However, it is only possible if 
$\g=0$ unless $\hat{J}^a\neq 0$. Hence, we have
$$ \g=0~,\quad {\rm or} \quad \hat{J}^m=\hat{J}^8=0~.\eqno(4.21)$$
Second, there are different unwanted terms of the form
$$ 2\b (C_{mpt} \hat{J}^p \j^t + \hat{J}^m \j^8 ) \j^r + (m\leftrightarrow r)~,
\eqno(4.22a)$$
in the OPE for $S^m(z) S^r(w)$, and that of the form
$$ 2\b (C_{rnp} \j^8 \j^p + \j^n \j^r ) \hat{J}^n ~,\eqno(4.22b)$$
in the OPE for $S^8(z) S^r(w)$. These unwanted terms vanish if and only if
$$ \b=0~,\quad {\rm or} \quad \hat{J}^m=0~.\eqno(4.23)$$
To this end, we examine in detail  both non-trivial possibilities:\\
(i) ~~$\hat{J}^m\neq 0~, \quad{\rm and}\quad \b=\g=0$, \\
(ii) ~$\hat{J}^m=\g\hat{J}^8=0~, \quad{\rm and}\quad \b\neq 0$.
\vglue.2in

{\bf (i)}. This case corresponds to having no trilinear fermions in the Ansatz
$(\b=0)$, as well as no background charge $(\g=0)$. From eq.~(4.7b) we find
$$\eqalign{
\fracmm{i(k+2)}{6(k+4)}T^{mn}=~&~\a^2\left\{ -\hat{J}^{mn} -4\hat{k}\j^m\j^n
+4\hat{k}C\ud{mn}{p}\j^p\j^8 
+\ha(\hat{k}-\frac{1}{8}\hat{k}_1)C\ud{mn}{pq}\j^p\j^q \right\}~,\cr
\fracmm{i(k+2)}{6(k+4)}C\ud{m}{pq}T^{pq}=~&~\a^2\left\{ -C^m_{~pq}\hat{J}^{pq} 
 + 4\hat{k}C\ud{m}{pq}\j^p\j^q + \left( 4\hat{k} + \ha\hat{k}_1\right)\j^m\j^8 
\right\}~.\cr} \eqno(4.24)$$
Multiplying the first line of this equation by $C\ud{p}{mn}$ and comparing the
result with the second line yields two equations for the coefficients at
the fermionic terms. Fortunately, these two equations turn out to be the same if we set:
$$\hat{k}_1=40\hat{k}~.\eqno(4.25)$$

The first line of eq.~(4.24) now takes the form
$$T^{mn} = \fracmm{i6(k+4)}{k+2}\a^2\left\{\hat{J}^{mn} 
+ 2\hat{k}\left(2\j^m\j^n -2C\ud{mn}{p}\j^p\j^8
+C\ud{mn}{pq}\j^p\j^q\right)\right\}~,\eqno(4.26)$$
where we can recognize the fermionic $Spin(7)$ generators, because of 
the identity
$$2\j^{[m}\j^{n]} + C\ud{mn}{pq}\j^p\j^q -2C\ud{mn}{p}\j^p\j^8=\bar{\j}
\tilde{\g}^{mn}\j~, \eqno(4.27)$$
in terms of the Majorana spinor $\j$, $\bar{\j}=\j^{\dg}=\j^{\rm T}$, with the
 components $\j^a$, if we take the lower sign in eq.~(2.14). Note also the 
related identities
$$\eqalign{
C\ud{p}{mn}\bar{\j}\tilde{\g}^{mn}\j= ~&~ 
-2\left(C\ud{p}{mn}\j^m\j^n + 6\j^p\j^8\right)~, \cr
C\ud{pq}{mn}\bar{\j}\tilde{\g}^{mn}\j= ~&~ 
8\left(\j^p\j^q + C\ud{pq}{m}\j^m\j^8\right)~. \cr}\eqno(4.28)$$

Eq.~(4.7a) consistently produces $\d^{MN}$ on the r.h.s. (the identities 
(4.18) play an important role here!) and determines $\a$,
$$\a^2 = -\fracmm{k(k+2)}{36\hat{k}(k+4)}~,\eqno(4.29)$$
while eq.~(4.26) can now be rewritten in an explicitly $SO(7)$-covariant form,
$$T^{mn}(z)=-\fracmm{ik}{6\hat{k}}\left\{\hat{J}^{mn}(z) +
2\hat{k}:\bar{\j}\tilde{\g}^{mn}\j:(z) \right\}~.\eqno(4.30)$$

Eq.~(4.30) determines the level $k$ of the $N=8$ algebra, by comparing 
the double-pole contributions on the both sides of the OPE defining the 
$\Hat{SO(7)}$ algebra that these currents satisfy. A direct calculation gives
$$ k = \fracmm{k^2}{9\hat{k}^2}\left( \hat{k}+4\hat{k}^2\cdot 1\right)~, \quad
{\rm or} \quad k=\fracmm{9\hat{k}}{1+4\hat{k}}~.\eqno(4.31)$$
 As a result, we get the following simple expression for the $Spin(7)$ 
current $T^{mn}$:
$$ T^{mn}(z)= -\,\fracmm{3i}{2(1+4\hat{k})}\left\{ \hat{J}^{mn}(z)+ 2\hat{k}
:\bar{\psi}\tilde{\gamma}^{mn} \psi : (z) \right\} ~.\eqno(4.32) $$

Taking the trace in eq.~(4.7c) with respect to the indices $M=N$, and using
$\d^{MM}=8$ and the obvious properties of the gamma matrices:  
$\tr(\g_{mn})= \tr(\g_{mnpq})=0$, we find
$$\eqalign{
T - \fracmm{1}{6(k+4)}:T^{mn}T^{mn}:\,  = ~&~
\a^2 \left[ - :\hat{J}^a\hat{J}^a: + \fracmm{1}{8}(7\tilde{k}+40\hat{k})
:\j^a\pa\j^a: - \hat{J}^{mn}:\j^m\j^n: \right. \cr
~&~\left. -\ha C_{mnpq}\hat{J}^{mn}:\j^p\j^q: +  C_{mnp}\hat{J}^{mn}:\j^p\j^8:
\,\right]\cr} \eqno(4.33)$$
where eqs.~(2.9), (2.11) and (4.17), the book-keeping definition 
$\hat{J}^a\equiv(\hat{J}^m,\hat{J}^8)$, as well as the identities of Appendix 
A, have been used. 

Next, making use of the definitions 
$$\eqalign{
\g^m_{ab} = i\left( C^m_{~ab} +\d^m_a\d^8_b - \d^m_b\d^8_a\right)~,\quad 
~&~ \quad \g^{mn}=\g^{[m}\g^{n]}~, \cr
\tilde{\g}^m_{ab} = i\left( C^m_{~ab} - \d^m_a\d^8_b + \d^m_b\d^8_a\right)~,
\quad ~&~ \quad \tilde{\g}^{mn}=\tilde{\g}^{[m}\tilde{\g}^{n]}~,\cr}
\eqno(4.34)$$
and the related identities (4.27) together with the identity
$$C_{mnpq}:\j^m\j^n\j^p\j^q: + 4C_{mnp}:\j^m\j^n\j^p\j^8: =
\ha :(\bar{\j}\tilde{\g}^{mn}\j)(\bar{\j}\tilde{\g}^{mn}\j):~,\eqno(4.35)$$
we can rewrite eq.~(4.33) in a compact and elegant form,
$$ T - \fracmm{1}{6(k+4)}:T^{mn}T^{mn}:\, =  
\a^2\left[ - :\hat{J}^a\hat{J}^a: + \fracmm{1}{8}(7\tilde{k}+40\hat{k})
:\bar{\j}\,\pa\j: -\ha\hat{J}^{mn}:(\bar{\j}\tilde{\g}^{mn}\j):\,\right]~.
\eqno(4.36)$$
Eq.~(4.36) is also explicitly $SO(7)$-covariant, which is important for the
consistency of our calculations. The coefficients in eq.~(4.36) follow from 
eqs.~(4.14), (4.29) and (4.31):
$$\tilde{k}=8\hat{k}~,\qquad 
\a^2=-\,\fracmm{2+17\hat{k}}{4(1+4\hat{k})(4+25\hat{k})}~~.\eqno(4.37)$$
Hence, we get from eq.~(4.36) that the $N=8$ stress tensor is given by
$$\eqalign{
T =~&~\fracmm{1}{8(1+\hat{k})(4+25\hat{k})} \left\{ 
- 3:\hat{J}^{mn}\hat{J}^{mn}: + 2(2+17\hat{k}):\hat{J}^a\hat{J}^a: 
-24\hat{k}(2+17\hat{k}):\bar{\j}\,\pa\j: \right. \cr
~&~\left. +(2+5\hat{k})\hat{J}^{mn}:(\bar{\j}\tilde{\g}^{mn}\j): -12\hat{k}^2 
:(\bar{\j}\tilde{\g}^{mn}\j)(\bar{\j}\tilde{\g}^{mn}\j):\,\right\}~.\cr}
\eqno(4.38)$$
where we have used eq.~(4.33). It is straightforward to check the rest of the
$N=8$ algebra OPEs. In particular, all the equations (4.7) now become 
 identities.

Since the level of an affine Lie algebra based on a compact Lie group must be
a positive integer for unitary highest-weight representations, eq.~(4.31) 
implies that we must consider either non-highest-weight-type unitary 
representations or non-unitary representations of $\Hat{SO(8)}$, in order to 
have a positive integer $k$, in general. The only exception exists when 
$\hat{k}=2$, which yields $k=2$ also. According to eq.~(3.5), the corresponding
central charge is given by 
$$c_8=10~.\eqno(4.39)$$ 
The full list of the $N=8$ algebra generators in the case of $\hat{k}=k=2$,
$c_8=10$, reads:
$$\eqalign{
T^{mn} = ~&~ - \fracmm{i}{6} \left\{ \hat{J}^{mn} + 4\,\bar{\psi}
 \tilde{\gamma}^{mn} \psi \right\}~,\cr
S^m =  ~&~ \fracmm{2i}{3\sqrt{6}}\,\g^m_{ab}\j^a\hat{J}^b~,\qquad
 S^8 =  \fracmm{2}{3\sqrt{6}}\, \j^a\hat{J}^a~, \cr
T   =   ~&~ \fracmm{1}{18}:\hat{J}^a\hat{J}^a: 
-\,\fracmm{1}{432}:\hat{J}^{mn}\hat{J}^{mn}: -\,\fracmm{4}{3}:\bar{\j}\,\pa\j:
\cr
&~ + \fracmm{1}{108}\hat{J}^{mn}:(\bar{\j}\tilde{\g}^{mn}\j):
 -\,\fracmm{1}{27}:(\bar{\j}\tilde{\g}^{mn}\j)(\bar{\j}\tilde{\g}^{mn}\j):~. 
\cr}\eqno(4.40)$$

We should note however that the choice $\hat{k}=2$ is not consistent with the
defining (anti)-commutation relations of the $N=8$ algebra since by  repeated
commutation of the current 
$T^{mn}\sim (\hat{J}^{mn} + 4 \bar{\psi} \tilde{\gamma}^{mn} \psi )$
with itself one generates currents of the form  
$(\hat{J}^{mn}+ 4^l \bar{\psi} \tilde{\gamma}^{mn} \psi )$, where 
$l=1,2,3,\ldots\,$. If we choose $\hat{k}=2$ we will have to extend the algebra
to a larger one.

Since the affine current $T^{mn}$ of the $N=8$ algebra in eq.~(4.30) is a 
linear combination of the bosonic and fermionic contributions, 
$-(i/2)\hat{J}^{mn}$ and $-(i/2)\bar{\psi}\tilde{\gamma}^{mn}\psi$, all having
the same (classical) normalisation, $T^{mn}$ would be {\it precisely} given by
their sum only if $k=\hat{k}+1$. This is consistent with eqs.~(4.30) and (4.31) 
if and only if $\hat{k}=1/2$ and $k=3/2$. The full list of the $N=8$ algebra 
generators in the case of $\hat{k}=1/2$, $k=3/2$ and $c_8=84/11$ is given by
$$\eqalign{
T^{mn} = ~&~ - \fracmm{i}{2} \left\{ \hat{J}^{mn} + \bar{\psi}
 \tilde{\gamma}^{mn} \psi \right\}~,\cr
S^m =  ~&~ i\sqrt{\fracmm{7}{33}}\,\g^m_{ab}\j^a\hat{J}^b~,\qquad
 S^8 =  \sqrt{\fracmm{7}{33}}\, \j^a\hat{J}^a~, \cr
T   =   ~&~ \fracmm{1}{132}\left\{ 7:\hat{J}^a\hat{J}^a:
-:\hat{J}^{mn}\hat{J}^{mn}: -42:\bar{\j}\,\pa\j:\right. \cr
&~\left. + \fracmm{3}{2}\hat{J}^{mn}:(\bar{\j}\tilde{\g}^{mn}\j):
 - :(\bar{\j}\tilde{\g}^{mn}\j)(\bar{\j}\tilde{\g}^{mn}\j):\,\right\}~.
\cr}\eqno(4.41)$$
Another consistent solution is to start from the {\it non-compact} real form 
$\Hat{SO(7,1)}$ in our Ansatz, and take its level to be $\hat{k}=-1/2$. Using
eq.~(4.31), this yields the level $k=9/2$ for the affine $SO(7)$ symmetry of
the $N=8$ algebra, and a central charge $c_8=360/17$ according to eq.~(3.5).

In both cases of consistent solutions the corresponding $(1,0)$ supersymmetric 
gauged WZNW models with the target space $SO(8)/SO(7) \times U(1)$ or 
$SO(7,1)/SO(7) \times U(1)$ must therefore have a hidden non-linear $N=8$ 
superconformal symmetry on-shell.

\vglue.2in

{\bf (ii)}. Because of eq.~(4.21), we are to distinguish the 
two possibilities: (a) 
without a $U(1)$ current $(\hat{J}^8=0)$ but with a background charge 
$(\g\neq 0)$, and (b) vice versa, $\hat{J}^8\neq 0$ but $\g=0$. 

The analogue of eq.~(4.24) in the case (ii), $\hat{J}^m=0$ and $\b\neq 0$, is
given by
$$
\fracmm{i(k+2)}{6(k+4)}T^{mn}=\a^2\left\{ 2\j^m\j^n [2\b^2(\Bar{B}^2 + 
\Bar{C}^2)-2i\b\g\Bar{B}]-2C\ud{mn}{p}\j^p\j^8[-2i\b\g\Bar{C}+4\b^2\Bar{B}\,
\Bar{C}]  \right.$$
$$ \left. ~~~~~~~~~~~~
+ C\ud{mn}{pq}\j^p\j^q [-\frac{1}{2}\b^2\Bar{C}^2 -\frac{1}{16}\hat{k}_1
 +i\b\g\Bar{B}\,] \right\} \eqno(4.42)$$
$$\fracmm{i(k+2)}{6(k+4)}C\ud{m}{pq}T^{pq}= \a^2\left\{ C\ud{m}{pq}\j^p\j^q 
[2i\b\g(\Bar{A}-\Bar{C}) -8\b^2\Bar{A}\,\Bar{B}] 
+\j^m\j^8[\ha\hat{k}_1 -12\b^2\Bar{A}\,\Bar{C}\,] \right\}~.$$
These two equations are only compatible if
$$\eqalign{
4\b^2\Bar{B}^2 +6\b^2\Bar{C}^2 +\frac{1}{4}\hat{k}_1 - 4i\b\g\Bar{B}~=~&~
 2i\b\g(\Bar{A}-\Bar{C})- 8\b^2\Bar{A}\,\Bar{B}~,\cr
-48i\b\g\Bar{C}+96\b^2\Bar{B}\,\Bar{C}~=~&~24\b^2\Bar{A}\,\Bar{C}-\hat{k}_1~,}
\eqno(4.43)$$
where we have to add the additional condition $\g\hat{k}_1=0$ from eq.~(4.21).

It is not difficult to check that there is only one consistent solution of 
eq.~(4.43) which is compatible with the $SO(7)$ symmetry, namely,
$$\g=0~,\qquad \Bar{A}=\Bar{B}=\Bar{C}=1~,\quad{\rm and}\quad 
\hat{k}_1=-72\b^2~.\eqno(4.44)$$
The first line of eq.~(4.42) thus takes the form
$$T^{mn}=-\fracmm{i}{2}k(\bar{\j}\tilde{\g}^{mn}\j)~,\eqno(4.45)$$
where we have used eq.~(4.27) and the eq.~(4.7a) gives :
$$\a^2 = \fracmm{k(k+2)}{48\b^2(k+4)}~.\eqno(4.46)$$  
Eqs.~(4.45) and (3.5) now imply that
$$ k=1~,\qquad {\rm and}\qquad c_8=26/5~,  \eqno(4.47)$$
respectively. The list of the $N=8$ superconformal algebra generators in the 
case (ii) is given by
$$\eqalign{
 T^{mn} = ~&~ - \fracmm{i}{2} \left( \bar{\psi}
 \tilde{\gamma}^{mn} \psi \right)~,\cr
 S^m =  ~&~ \fracmm{i}{\sqrt{5}} \left( i\j^m\hat{J}^8 +\fracmm{1}{3}
 C\ud{m}{npq}\j^n\j^p\j^q + C\ud{m}{pq}\j^p\j^q\j^8\right)~,\cr
 S^8 =  ~&~ \fracmm{i}{\sqrt{5}}\left( -i\j^8\hat{J}^8 
+ C_{mnp}\j^m\j^n\j^p \right)~, \cr
 T   =   ~&~ \fracmm{1}{20}:\hat{J}^8\hat{J}^8:-\fracmm{3}{8}:\bar{\j}\,\pa\j: 
+\fracmm{1}{240}:(\bar{\j}\tilde{\g}^{mn}\j)(\bar{\j}\tilde{\g}^{mn}\j):~.\cr}
\eqno(4.48) $$
This case thus corresponds to a unitary realization of the $N=8$ algebra in 
terms of a single free boson and eight free fermions transforming in 
${\underline{1}}$ and ${\underline{8}}$ of $Spin(7)$, respectively.  
\vglue.2in

\subsection{A construction of the exceptional ~$N=7$~ superconformal algebra
                  over the coset space ~$SO(7)\times U(1)/G_2$}

The $N=7$ coset construction over $SO(7)\times U(1)/G_2$ follows the lines 
of the $N=8$ case considered above. In the $N=7$ case, our starting 
point is the affine algebra $\Hat{SO(7)}_{\hat{k}}$ whose commutation relations
can be read off from  eq.~(4.12) by restricting the indices to run from one to
seven ($m,n,\ldots=1,\ldots,7$). Eqs.~(2.17) and (2.20) imply the following 
definitions of the currents associated with the coset $SO(7)/G_2$ and the group
 $G_2$, respectively,
$$\hat{A}^m(z) = \ha C\ud{m}{np}\hat{J}^{np}(z)~,\quad {\rm and} \quad
\hat{G}^{mn}(z)\equiv \ha \hat{J}^{mn}(z) 
+ \fracm{1}{8}C\ud{mn}{pq}\hat{J}^{pq}(z)~.\eqno(4.49)$$
Accordingly, we get the OPE
$$ \hat{A}^m(z) \hat{A}^n(w) = \fracmm{-12\hat{k}}{(z-w)^2}\d^{mn} +
\fracmm{2C\ud{mn}{k}\hat{A}^k -8\hat{G}^{mn}}{z-w} + :\hat{A}^m\hat{A}^n:(w) 
+\ldots~.\eqno(4.50)$$

We shall denote the affine factor $\Hat{U(1)}$ by a bosonic current 
$\hat{A}^0(z)$, with the (normalised) OPE ({\it cf}. eq.~(4.13))
$$\hat{A}^0(z)\hat{A}^0(w) = \fracmm{1/2}{(z-w)^2} + :\hat{A}^0\hat{A}^0:(w)
+\ldots~,\eqno(4.51)$$
and define the 8=1+7 free fermions $\j^a(z)$ to be $(\j^m,\j^8)$, with the OPE 
as in eq.~(4.4).

Our Ansatz for the supercurrents of the $N=7$ non-linear superconformal algebra
 reads as follows:
$$\eqalign{
S^m =~&~ 2\a \left\{ C\ud{m}{np}\j^n\hat{A}^p + a\j^8\hat{A}^m + b\j^m\hat{A}^0
+\g\pa\j^m \right. \cr
~&~~~ \left.  -\fracmm{2i}{3}\b \left[ C\ud{m}{npq}\j^n\j^p\j^q 
 +3d C\ud{m}{np}\j^8\j^n\j^p\right] \right\} ~,\cr} \eqno(4.52)$$
where $\a,\b,a,b$ and $d$ are  parameters to be 
determined by the OPEs of the $N=7$ algebra.

In terms of the general Ansatz (4.5), eq.~(4.52) implies that we equate
$$h^m_{np} =C\ud{m}{np}~,\quad h^m_{8n} =a\d^m_n~,\quad h^m_{n8} =b\d^m_n~;
\qquad  \x^m_n=\g\d^m_n~, \eqno(4.53)$$
and 
$$\G^m_{npq} =C\ud{m}{npq}~,\qquad \G^m_{np8} =dC\ud{m}{np}~.\eqno(4.54)$$
It follows~\footnote{Note our conventions: the {\it early} lower-case 
Latin indices take values $a,b,\ldots =1,2,\ldots,8$, whereas the {\it middle}
 lower-case Latin indices take values $i,j,\ldots =1,2,\ldots,7$.}
$$\eqalign{
h^m_{ab}h^n_{ab} = & \left( 6 + a^2 + b^2\right) \d^{mn}~,\cr
h^m_{ab}h^n_{ad} = & h^m_{pb}h^n_{pd} + h^m_{8b}h^n_{8d}~,\cr}\eqno(4.55)$$
where
$$\eqalign{
h^m_{aq}h^n_{ar}= & C_{mnqr} -\d_{qn}\d_{mr} +\d_{mn}\d_{qr}
+a^2\d_{mq}\d_{nr}~,\cr
h^m_{a8}h^n_{a8} = & b^2\d^{mn}~,\cr
h^m_{a8}h^n_{ap}= & -bC_{mnp}~,\cr
h^m_{ap}h^n_{a8}= & +bC_{mnp}~,\cr }\eqno(4.56)$$
and, similarly,
$$\eqalign{
h^m_{pc}h^n_{qc}= & C_{mnpq} +\d_{mn}\d_{pq} - \d_{mq}\d_{pn}
 +b^2\d_{mp}\d_{nq}~,\cr
h^m_{8c}h^n_{8c} = & a^2\d^{mn}~,\cr
h^m_{8c}h^n_{pc}= & +a C_{mnp}~,\cr
h^m_{pc}h^n_{8c}= & -a C_{mnp}~.\cr}\eqno(4.57)$$

In addition, we find
$$\eqalign{
\G^m_{abc}\G^n_{abc} = & 6\left(4+3d^2 \right)\d^{mn}~,\cr
\G^m_{abc}\G^n_{abd} = & \G^m_{pqc}\G^n_{pqd} + 2 \G^m_{8pc}\G^n_{8pd}~,\cr}
\eqno(4.58)$$
where
$$\eqalign{
\G^m_{abr}\G^n_{abs} = & 2\left(1+d^2\right)C\ud{mn}{rs} +
2\left(2+d^2\right)\left(\d^m_n\d^r_s - \d^m_s\d^r_n\right)~,\cr  
\G^m_{abr}\G^n_{ab8} = & -4dC^{mnr}~,\cr
\G^m_{ab8}\G^n_{abr} = & +4dC^{mnr}~.\cr}\eqno(4.59)$$
Some other useful corollaries of eq.~(4.52) are
$$\eqalign{
\fracm{i}{2}h^m_{ab}h^n_{ad}f^{bdE}\hat{J}^E =& -4\left(3+a^2\right)
\hat{G}^{mn} + \left(a^2-3\right)C^{mnp}\hat{A}^p~,\cr
h^m_{ab}h^n_{cb}\j^a\j^c=& \left(1+b^2\right)\j^m\j^n + C\ud{mn}{pq}\j^p\j^q
+2aC^{mnp}\j^8\j^p~,\cr
2\b^2 \G^m_{abc}\G^n_{abd}\j^c\j^d = & 4\b^2\left( 2+d^2 \right)\j^m\j^n
 +  4\b^2\left(1+d^2\right)C\ud{mn}{pq}\j^p\j^q \cr
{} & + 16\b^2 dC^{mnp}\j^8\j^p~.\cr} \eqno(4.60)$$

We are now in a position to calculate the r.h.s. of eqs.~(4.8b) and (4.8a), by
using the relation $\tilde{k}=-24\hat{k}$ which follows from eq.~(4.50), and 
the equations above. We find
$$\eqalign{
\fracmm{3k+5}{k+3} (M^A)^{mn}G^A = ~&~ 
4\a^2 \left\{ -4(3+a^2)\hat{G}^{mn} + (a^2-3+\g)C^{mnp}\hat{A}^p \right. \cr
{} ~&~ + 4\left[ \b^2(2+d^2) - 3\hat{k} +\frac{1}{8} b^2 \right]\j^m\j^n \cr
{} ~&~ + \left[ 4\b^2(1+d^2) -12\hat{k}-\frac{4i}{3}\b\g\right] 
C\ud{mn}{pq}\j^p\j^q \cr
{} ~&~ \left. +4 \left[ 4\b^2 d -6\hat{k}a + i\g\b d\right] 
C^{mnp}\j^8\j^p\right\}~, \cr} \eqno(4.61)$$
and 
$$ \fracmm{k(3k+5)}{k+3}=4\a^2\left[ 2\b^2(4+3d^2) -6\hat{k}(6+a^2)
+\frac{1}{4}b^2 -\g^2 \right]~.\eqno(4.62)$$

The stress tensor $T$ of the $N=7$ algebra can be calculated from the OPE of 
the given supercurrents $S^m$ in eq.~(4.52) along the lines of the previous 
subsection. Taking the trace in eq.~(3.11) and using eq.~(3.6), we find 
$$S^m(z)S^m(w)~\sim~\fracmm{7k(3k+5)}{(k+3)(z-w)^3} +
\fracmm{1}{z-w}\left[ 14T(w)+\fracmm{4}{3(k+3)}:G^{mn}G^{mn}:(w)\right]~.
\eqno(4.63)$$
Since $\tr(M^A)=0$ and $\tr(M^AM^B)=2\d^{AB}$, eqs.~(4.8c) and (4.63) imply
$$ 7T - \fracmm{3}{\hat{k}+4}:G^AG^A:\,= \a^2 \left\{  h^m_{ab} h^m_{ad}
:\hat{A}^b\hat{A}^d: -\tilde{k}h^m_{ab} h^m_{cb}:\j^c\pa\j^a:\right.$$
$$ + 4  h^m_{ap} h^m_{cq}(C_s^{pq}\hat{A}^s -4\hat{G}^{pq}):\j^a\j^c: -
4i\b \G^m_{abc}h^m_{cf}\hat{A}^f:\j^a\j^b: 
 -4\b^2\G^m_{abf}\G^m_{abg}:\j^g\pa\j^f: $$
$$\left. +7\g b\pa\hat{A}^0
-4\b^2\G^m_{abc}\G^m_{afg}:\j^b\j^c\j^f\j^g:\right\}~,\eqno(4.64)$$
where $m,n,\ldots=1,\ldots,7$, and  $a,b,\ldots=1,\ldots,8$.

In accordance with our definitions, we have
$$\eqalign{
 h^m_{ab} h^m_{ad}:\hat{A}^b\hat{A}^d:\,=~&~ (6+a^2):\hat{A}^m\hat{A}^m:
+7b^2:\hat{A}^0\hat{A}^0:~,\cr
h^m_{ab} h^m_{cb}:\j^c\pa\j^a:\,=~&~ (6+b^2):\j^m\pa\j^m: 
+ 7a^2:\j^8\pa\j^8:~, \cr
h^m_{ap} h^m_{cq}(C\ud{pq}{s}\hat{A}^s -4\hat{G}^{pq}):\j^a\j^c:\,=~&~ 
12a\hat{A}^m:\j^m\j^8: -3(C\ud{mn}{p}\hat{A}^p +4\hat{G}^{mn}):\j^m\j^n:~,\cr
\G^m_{abc}h^m_{cf}\hat{A}^f:\j^a\j^b:\,=~&~
12d\hat{A}^m:\j^m\j^8: + (a-4)C\ud{mn}{p}\hat{A}^p:\j^m\j^n:~,\cr
\G^m_{abf}\G^m_{abg}:\j^g\pa\j^f: \,=~&~
12(2+d^2):\j^m\pa\j^m: +42d^2:\j^8\pa\j^8:~,\cr}\eqno(4.65)$$
and
$$\G^m_{abc}\G^m_{afg}:\j^b\j^c\j^f\j^g:\,=
-(2+d^2)C_{mnpq}:\j^m\j^n\j^p\j^q:
-16 d :\j^m\j^n\j^p\j^8:~.\eqno(4.66)$$
Hence, we can rewrite eq.~(4.64) as follows:
$$  7T - \fracmm{3}{\hat{k}+4}:G^AG^A:\,=  \a^2 \left\{ 
(6+a^2):\hat{A}^m\hat{A}^m: + 7b^2:\hat{A}^0\hat{A}^0:
+7\g b\pa\hat{A}^0\right.$$
$$ + \left[ 144\hat{k}-48\b^2(2+d^2)-b^2\right] :\j^m\pa\j^m: 
 + 168(\hat{k}a^2-\b^2d^2) :\j^8\pa\j^8: $$
$$ -48\hat{G}^{mn} :\j^m\j^m: +4\left[ i\b(4-a)-3\right]
C_{pmn}\hat{A}^p :\j^m\j^n:  +48(a-i\b d)\hat{A}^m :\j^m\j^8: $$
$$ \left. + 4\b^2(2+d^2)C_{mnpq}:\j^m\j^n\j^p\j^q: 
+64\b^2 dC_{mnp}:\j^m\j^n\j^p\j^8:\right\}~.\eqno(4.67)$$

The terms bilinear or quartic in the fermionic fields $\j^m$ have to
appear in the currents of the $N=7$ algebra in a covariant form, i.e. in the
$G_2$-covariant combination 
$$\bar{\j}g^{mn}\j \equiv \bar{\j}\left( \ha\g^{mn}+\fracm{1}{8}C\ud{mn}{pq}
\g^{pq}\right)\j = 2\j^m\j^n + \ha C\ud{mn}{pq}\j^p\j^q~.\eqno(4.68)$$
Some related identities are given by
$$\eqalign{
\hat{G}^{mn}:\j^m\j^n:\,=~&~ \fracmm{1}{3}\hat{G}^{mn}:(\bar{\j}g^{mn}\j)~,\cr
C^p_{mn}\hat{A}^p :\j^m\j^n:\,=~&~ -i\hat{A}^m:(\bar{\j}\tilde{\g}^m\j): +
2\hat{A}^m:\j^m\j^8:~,\cr
C_{mnp}:\j^m\j^n\j^p\j^8:\,=~&~ -i:(\bar{\j}\tilde{\g}^m\j) \j^m\j^8:~,\cr
C_{mnpq}:\j^m\j^n\j^p\j^q:\,=~&~\fracmm{2}{3}:(\bar{\j}g^{mn}\j)
(\bar{\j}g^{mn}\j):~.\cr}\eqno(4.69)$$
When using these identities and eqs.~(2.11) and (2.19), we arrive at the 
explicitly $G_2$-covariant expression for the stress tensor, namely,
$$T = \fracmm{3}{7(\hat{k}+4)}:G^AG^A:\,+ \fracmm{\a^2}{7} \left\{
(6+a^2):\hat{A}^m\hat{A}^m: + 7b^2:\hat{A}^0\hat{A}^0: 
+7\g b\pa\hat{A}^0   \right.$$
$$ + \left[ 144\hat{k} -48\b^2(2+d^2) - b^2\right] :\j^m\pa\j^m: 
 + 168(\hat{k}a^2-\b^2d^2) :\j^8\pa\j^8: $$
$$ +4\left[ \b(4-a)+3i\right]\hat{A}^m :(\bar{\j}\tilde{\g}^m\j):
+8\left[6(a-i\b d)+i\b(4-a)-3\right]\hat{A}^m :\j^m\j^8: $$ 
$$\left.  -16\hat{G}^{mn} :(\bar{\j}g^{mn}\j):
 -64i\b^2 d:(\bar{\j}\tilde{\g}^m\j)\j^m\j^8: +\fracm{8}{3}\b^2(2+d^2)
:(\bar{\j}g^{mn}\j)^2:\right\}~,\eqno(4.70)$$
where $\hat{A}^m$, $\j^m$ and $:(\bar{\j}\tilde{\g}^m\j):$ all transform in 
${\underline{7}}$ of $G_2$. 

Once all the currents of the $N=7$ algebra are determined, we are to consider
the consistency conditions which follow from eq.~(4.61) and (4.8c).  The 
terms in (4.8c) that are symmetric and traceless with  respect to  $(m,n)$ indices 
determine the composite field $:GG:$ which should not depend on the coset
currents $\hat{A}^{m}$. Similarly to the $N=8$ case, it is not
difficult to check that this is only possible if either \\
(i)~ $\hat{A}^m\neq 0$, $\g\neq 0$ and $\b=d=0$, \\
or \\
(ii)~ $\hat{A}^m=0$, $\g=0$ and $\b\neq 0$.\\
Each case is separately considered below.
\vglue.2in

{\bf (i)}. It follows from eqs.~(4.61) and (4.68) that 
$$ a^2+\g=3~,\qquad a\hat{k}=0~,\quad b^2+72\hat{k}=0~.\eqno(4.71)$$
Therefore, in order to have a non-trivial solution $(\hat{k}\neq 0)$, we must
take 
$$ a=0~,\qquad \g=3~,\qquad b^2=-72\hat{k}~.\eqno(4.72)$$ 
Eqs.~(4.61) and (4.62) now imply ({\it cf}. eq.~(4.30))
$$G^A = -\,\fracmm{1}{3}\left(M^A\right)^{mn}\left[ \fracmm{2k}{1+6\hat{k}}
\right] \left\{ \hat{G}^{mn} + 2\hat{k}\left(\bar{\j}g^{mn}\j\right)\right\}~,
\eqno(4.73)$$
and
$$ \fracmm{k(3k+5)}{k+3}=-36\a^2(1+6\hat{k})~.\eqno(4.74)$$

Since the relative normalisation of the $G_2$ generators $\hat{G}^{mn}$ and
$\left(\bar{\j}g^{mn}\j\right)$ is the same, it follows from eq.~(4.73) that
$2\hat{k}=1$, which implies $k=\hat{k}+1=3/2$, exactly as in the $N=8$ case~!
This is also consistent with the level equation associated with eq.~(4.73).
The affine currents of the $N=7$ algebra now take the very simple form:
$$G^A(z) = -\fracmm{1}{4}(M^A)^{mn}
\left\{ \hat{G}^{mn} + \left(\bar{\j}g^{mn}\j\right)\right\}~,\eqno(4.75)$$
which is quite similar to the expression for the $N=8$ affine currents in 
eq.~(4.41). Eqs.~(3.6) and (4.75) also imply
$$\eqalign{ 
:G^AG^A:\,=~&~ -\fracmm{1}{8}:\left(\hat{G}^{mn}+\bar{\j}g^{mn}\j\right)^2: 
\cr
{}~&~ = -\fracmm{1}{8}:\left[ :\hat{G}^{mn}\hat{G}^{mn}: 
+2\hat{G}^{mn}:(\bar{\j}g^{mn}\j): + :(\bar{\j}g^{mn}\j)^2:\right]~.}
\eqno(4.76)$$ 

The remaining coefficient values are therefore given by $\a=(i/12)\sqrt{19/6}$
and $b=i6$. Eq.~(3.12) for $k=3/2$ yields $c_7=89/12$. The list of the $N=7$ 
currents in this realization reads as follows:
$$\eqalign{
G^A = ~&~ -\fracmm{1}{4}(M^A)^{mn}\left\{ 
\hat{G}^{mn}+\bar{\j}g^{mn}\j\right\}~, \cr
S^m  = ~&~ \fracmm{i}{6}\sqrt{\fracmm{19}{6}} \left\{ C\ud{m}{np}\j^n\hat{A}^p
 + i6\j^m\hat{A}^0 +3\pa\j^m \right\} ~,\cr
T = ~&~ -\fracmm{1}{84}:\left(\hat{G}^{mn} +\bar{\j}g^{mn}\j\right)^2: 
- \,\fracmm{19}{7\cdot 12^2} \left\{ :\hat{A}^m\hat{A}^m:
 -42 :\hat{A}^0\hat{A}^0:  +i21\pa\hat{A}^0 \right. \cr
{} ~&~ \left. +  18 :\j^m\pa\j^m: -16\hat{G}^{mn}:(\bar{\j}g^{mn}\j): 
 +12i\hat{A}^m:(\bar{\j}\tilde{\g}^m\j):  \right\}~,\cr}
\eqno(4.77)$$
where $\j^8=0$. It should be noted that the currents $\hat{A}^m$ are 
anti-hermitian. This leads to the extra factors of $i$ in the expressions for 
$S^m$ and $T$ so as to make all the terms appearing in them hermitian, as 
required by unitarity. 
\vglue.2in

{\bf (ii)}. Another realization of the $N=7$ algebra is possible in terms 
of one real boson and seven real fermions at the level $k=1$. It corresponds
to the following solution of the consistency equations:
$$ a=d=\g=0~,\qquad  b=4\b~,\qquad  \a^2\b^2=1/24~.\eqno(4.78)$$ 
The list of the $N=7$ currents in this case of $c_7=5$ reads:
$$\eqalign{
G^A = ~&~ -\fracmm{1}{3}(M^A)^{mn}\left( \bar{\j}g^{mn}\j\right)~, \cr
S^m  = ~&~ \fracmm{2}{\sqrt{6}} \left\{ 2\j^m\hat{A}^0 
-\fracmm{i}{3}C\ud{m}{npq}\j^n\j^p\j^q \right\} ~,\cr
T = ~&~ \fracmm{2}{3} :\hat{A}^0\hat{A}^0: -\fracmm{2}{3}:\j^m\pa\j^m: 
+ \fracmm{1}{126}:(\bar{\j}g^{mn}\j)(\bar{\j}g^{mn}\j):~.\cr}\eqno(4.79)$$
It is straightforward to check the rest of the $N=7$ algebra.
\vglue.2in

\section{Conclusion}

The main results of our investigation are given in sect.~4 where we presented
explicit realizations of the $N=7$ and $N=8$ non-linear superconformal 
algebras using coset space methods. The constraints of the non-linear algebras
allowed very few realizations for each algebra in the compact case, as well as
an additional realization for the $N=8$ algebra in the non-compact case, 
within our general Ansatz over specific coset spaces. It may be possible to 
find other realizations by considering other more general coset spaces.

Our results could be  relevant for string theory in various ways.
The exceptional superconformal symmetries may arise as hidden symmetries 
in certain compactifications of superstring theories. For example, there exist
 octonionic soliton solutions to the low-energy heterotic string theory 
\cite{hs,gn,i}. The octonionic soliton of ref.~\cite{hs} is related to the 
Yang-Mills instanton in eight dimensions with $SO(7)$ gauge  group 
\cite{fny,dwn}. The  octonionic soliton of ref.~\cite{gn} is related to 
the seven-dimensional Yang-Mills instanton with the gauge group $G_2$, which
is a remarkable instanton in odd dimensions \cite{gn} (see also 
ref.~\cite{ip} for other examples ). The conformal field-theoretic 
formulations of these remarkable octonionic solitons of the heterotic string 
may involve the exceptional superconformal algebras.

As was shown in ref.~\cite{gnst}, the light-cone gauge actions of various
superstring theories in the Green-Schwarz formalism have $N=8$ supersymmetry.
Since they are also conformally invariant, this implies that they must be
invariant under some $N=8$ supersymmetric extension of the Virasoro algebra.
Later, the Green-Schwarz superstring was shown to have the $N=8$ soft algebra
\cite{nb,bcp,cp,hs} as part of its constraint algebra. Whether the 
Green-Schwarz superstring action has a hidden non-linear $N=8$ or $N=7$ 
supersymmetry of the type investigated above is an interesting open problem.

A compactification of eleven-dimensional supergravity on the 7-sphere $S^7$ is
known to give $N=8$ supergravity in four dimensions, with the $SO(8)$ gauge
invariance \cite{dnp,dn87}. Similarly, the cosets  $S^7\times S^1$ or
 $S^7\times S^1\times S^1$ could be used to compactify the  eleven-dimensional
supergravity down to three or two dimensions. Recently, it has been realized 
that an $S^1$-compactified eleven-dimensional supergravity appears to be the 
low-energy effective field theory of the strongly coupled type-IIA superstring
which, in  turn, is related to the $S^1$-compactified eleven-dimensional 
supermembrane \cite{t1,w1}. In general, massless non-Abelian gauge fields do 
not arise as Kaluza-Klein modes in consistent string compactifications. It is 
however possible that the 11-dimensional M-theory may have consistent 
compactification over manifolds with non-trivial isometry groups such as the 
seven-sphere. If that is the case, then the $N=8$ and $N=7$ superconformal 
algebras may be the hidden symmetries of compactifications involving the 
seven-sphere.

Interestingly enough, there exist Ricci-flat seven- and eight-dimensional 
compact manifolds with the holonomy groups $G_2$ and $Spin(7)$, respectively 
\cite{j}. Both the ten-dimensional superstrings  and the eleven-dimensional 
supergravity compactified down to two and  three dimensions ( and  three and four 
dimensions) on such manifolds
have the minimal number (one) of supersymmetries in the corresponding 
dimension~\cite{sv,pt11}. Whether or not the exceptional superconformal algebras
can be  related to these constructions remains  to be investigated. 

Having obtained the unitary realizations of the exceptional superconformal
algebras with the central charges $c=26/5$ and $c=5$, one can use them to 
represent the $N=0$ and $N=1$ superconformal matter. Then, by `tensoring' five 
conformal matter models of 
$c=26/5$ and adding the conformal ghosts $(b,c)$, one gets an anomaly-free 
string model $(c_{\rm gh}=-26)$. Similarly,  by `tensoring' three 
superconformal matter models of $c=5$ and adding both the conformal and 
superconformal ghosts $(b,c)$ and $(\b,\g)$, one gets an anomaly-free 
superstring model $(c_{\rm gh}=-26+11=-15)$. Thus the
exceptional superconformal algebras may describe 
`exceptional' compactifications  of existing superstring theories and underlie
some novel `exceptional' strings and 
superstrings~!
\vglue.2in

\noindent{\Large\bf Acknowledgements}

We would like to thank J. Fuchs, O. Lechtenfeld, D. L\"ust, H. Nicolai, 
J. L. Petersen, A. Popov, C. Preitschopf and E. Sezgin for useful discussions. 
One of the authors (S.V.K.) wishes to acknowledge the hospitality of the
Department of Physics, Penn State University, extended to him during the 
initial period of this work.

\newpage

\noindent{\Large\bf Appendix A: Identities for the octonionic structure 
constants $C^{mnp}$ and $C^{mnpq}$}

The identities collected below follow from the definitions of tensors 
$C^{mnp}$ and $C^{mnpq}$ in subsect.~2.1. We find ({\it cf\/} 
refs.~\cite{wn,cdfn})\\
$$\eqalign{
C\ud{p}{mn}C\ud{mn}{q} = ~&~ 6\,\d^p_q~,\qquad C^{mnp}C^{mnp}=42~,\cr
C\ud{[q}{[mn}C\ud{ts]}{p]}= ~&~ -2C\ud{[qt}{[mn}\d^{s]}_{p]}~,\cr
C\ud{p}{mk}C\ud{q}{kn}= ~&~ \ha C\ud{pq}{k}C_{knm} - \frac{3}{2}C\ud{pq}{mn} 
+\ha\left(\d^p_m\d^q_n + \d^q_m\d^p_n\right) - \d^{pq}\d_{mn} ~;\cr}\eqno(A.1)
$$
$$\eqalign{
C_{mnkl}C\ud{kl}{p}= ~&~ -4\, C_{mnp}~,\cr
2C\ud{k}{[mn}C\ud{ks}{pq]}= ~&~ C\ud{k}{[mn}C\ud{ks}{p]q}
-C\ud{k}{q[m}C\ud{ks}{np]}~,\cr
C\ud{k[s}{[m}C\ud{t]k}{np]}=~&~ C\ud{[t}{[mn}\d^{s]}_{p]}~,\cr
C\ud{ks}{[m}C\ud{tk}{np]}=~&~C_{mnp}\d^{st}-C\ud{s}{[mn}\d^t_{p]}-2C\ud{t}{[mn}
\d^s_{p]}~,\cr
C\ud{k}{[mn}C\ud{pq}{s]k}=~&~4\, C\ud{[p}{[mn}\d^{q]}_{s]}~,\cr
C_{mnk}C^{kpqs}=~&~6\, C\ud{[pq}{[m}\d^{s]}_{n]}~,\cr
C\ud{tk}{[mn}C\ud{k}{pq]}=~&~-2\, C_{[mnp}\d^t_{q]}~,\cr
C_{[mnp}\d^s_{q]}-C_{[mnp}\d^q_{s]}=~&~\frac{3}{2}C\ud{[s}{[mn}\d^{q]}_{p]}~;
\cr}\eqno(A.2)$$
and
$$\eqalign{
C_{mnpk}C^{kqst}=~&~
9\, C\ud{[qs}{[mn}\d^{t]}_{p]} + 6\, \d^{[q}_{[m}\d^s_n\d^{t]}_{p]}~,\cr
C_{mnpq}C\ud{pq}{st}=~&~ -2\, C_{mnst}+4\, 
\left(\d_{ms}\d_{nt}-\d_{mt}\d_{ns}\right)~,\cr
C_{mkpq}C_{nkpq}=~&~24\, \d_{mn}~,\cr
C\ud{sk}{[mn}C\ud{tk}{pq]}=~&~ -\d^{st}C_{mnpq} - C\ud{s}{[mn}C\ud{t}{pq]}
+ 2C_{[mnp}{}^s\d^t_{q]} +  2C_{[mnp}{}^t\d^s_{q]}~.\cr}\eqno(A.3)$$
\vglue.2in

\newpage

\noindent{\Large\bf Appendix B: ~$N{=}8$~ supersymmetry algebra}

In this appendix, the ~$N{=}8$~ supersymmetry algebra of eq.~(3.4) is written
down in a more explicit form, after using the decomposition $\ul{8}=\ul{1}+
\ul{7}$ for the SO(7) spinors $S^a=(S^m,S^8)$, and substituting the gamma 
matrices in terms of the octonionic structure constants, as in sect.~2. We find
$$\eqalign{
S^8(z)S^8(w)~\sim~&~\fracmm{8k(k+2)}{3(k+4)}\fracmm{1}{(z-w)^3} 
+ \fracmm{2T(w)}{z-w}\cr
{} ~&~ - \fracmm{1}{3(k+4)}\fracmm{1}{(z-w)}\left\{ 
:T^{pq}T^{pq}: +\fracm{1}{4}C_{mnpq}:T^{mn}T^{pq}:\right\}(w) \cr
{} ~&~ = \fracmm{8k(k+2)}{3(k+4)}\fracmm{1}{(z-w)^3} +  \fracmm{2T(w)}{z-w} \cr
{}~&~ +\fracmm{1}{(k+4)}\fracmm{1}{(z-w)}\left\{ \fracm{1}{3} :A^pA^p:
 - \fracm{1}{2}:T^{pq}T^{pq}: \right\}(w)~,\cr}\eqno(B.1)$$
$$\eqalign{
S^m(z)S^m(w)~\stackrel{{\rm no~sum~over~}m}{\sim}
~&~\fracmm{8k(k+2)}{3(k+4)}\fracmm{1}{(z-w)^3} + \fracmm{2T(w)}{z-w} 
 - \fracmm{1}{3(k+4)}\fracmm{1}{(z-w)}\cr
{}~&~ \times \left\{ :T^{pq}T^{pq}: + \fracm{1}{4}C_{rspq}:T^{rs}T^{pq}: 
+ 2C_{mnpq}:T^{pq}T^{nm}: \right\}(w) \cr
{} ~&~ = \fracmm{8k(k+2)}{3(k+4)}\fracmm{1}{(z-w)^3} +  \fracmm{2T(w)}{z-w}\cr
{}~&~ +\fracmm{1}{(k+4)}\fracmm{1}{(z-w)}\left\{ \fracm{1}{3}:A^pA^p: -
 \fracm{1}{2}:T^{pq}T^{pq}: \right\}(w) \cr
{}~&~ +\fracmm{8}{9(k+4)}\fracmm{1}{(z-w)}\left\{ C^{mpq}:A_qT^{pm}: 
 -2:G^{mq}T^{qm}: \right\}(w)~,\cr}\eqno(B.2)$$
$$\eqalign{
S^m(z)S^8(w)~\sim~&~ \fracmm{i2(k+2)}{3(k+4)}C^m_{~pq}\left[ \fracmm{T^{pq}
(w)}{(z-w)^2} +\ha \fracmm{\pa T^{pq}(w)}{z-w} \right] \cr
{}~&~ +\fracmm{1}{3(k+4)}C_{npq}\fracmm{:T^{pq}T^{nm}:(w)}{(z-w)}\cr
{}~&~ = \fracmm{i4(k+2)}{3(k+4)}\left[ \fracmm{A^m(w)}{(z-w)^2} +\ha
 \fracmm{\pa A^m(w)}{(z-w)}\right]\cr
{}~&~ + \fracmm{2}{3(k+4)} \fracmm{:A_pT^{pm}:(w)}{z-w}~,\cr}\eqno(B.3)$$
and
$$\eqalign{
S^m(z)S^n(w)~\stackrel{m\neq n}{\sim}~&~\fracmm{i2(k+2)}{3(k+4)}\left[ 
C\ud{mn}{pq} + 2\d^m_p\d^n_q\right]\left(\fracmm{T^{pq}(w)}{(z-w)^2} 
+\ha \fracmm{\pa T^{pq}(w)}{z-w}\right)\cr
{}~&~ -\fracmm{1}{3(k+4)}\fracmm{1}{(z-w)}\left[ C\ud{ms}{pq}:T^{pq}T^{sn}: +
C\ud{ns}{pq}:T^{pq}T^{sm}:\right](w)\cr
 = ~&~  \fracmm{i4(k+2)}{9(k+4)}\left\{ \fracmm{8G^{mn}(w)
-C\ud{mn}{p}A^p(w)}{(z-w)^2} 
+ \fracmm{4\pa G^{mn}(w)-\frac{1}{2}C\ud{mn}{p} \pa A^p(w)}{z-w} \right\} \cr
{}~&~ - \fracmm{4}{9(k+4)}\fracmm{1}{(z-w)}\left[2:G^{ms}T^{sn}:(w)+
2:G^{ns}T^{sm}:(w)\right. \cr
{}~&~ \left. - C\ud{ms}{p} :A^pT^{sn}:(w) - C\ud{ns}{p} :A^pT^{sm}:(w)\right]~, 
\cr}\eqno(B.4)$$
where we have extensively used the identities from Appendix A. Note also that 
in this appendix the quantity $A^m$ represents $\ha C\ud{m}{np}T^{np}$.
\vglue.2in


\newpage

\begin{thebibliography}{99}

\bibitem{gsw} M. B. Green, J. H. Schwarz and E. Witten, {\it Superstring 
Theory}, Cambridge: Cambridge Univ. Press, 1987. 
\bibitem{jb} J. Fuchs, {\it Affine Lie Algebras and Quantum Groups},
Cambridge: Cambridge Univ. Press, 1992. 
\bibitem{book} S. V. Ketov, {\it Conformal Field Theory}, Singapore: World
Scientific, 1995.
\bibitem{gst} M. G\"{u}naydin, G. Sierra and P. K. Townsend, \np{274}{86}{429}.
\bibitem{rs} P. Ramond and J. H. Schwarz, \pl{64}{76}{75}.
\bibitem{af} L. Alvarez-Gaum\'e and D. Z. Freedman, \pl{94}{80}{171};\\
 \pr{22}{80}{846}; \cmp{80}{81}{443}.
\bibitem{sstp} P. Spindel, A. Sevrin, W. Troost and A. van Proeyen,
\np{308}{88}{662}, \ibid{311}{88}{465}.
\bibitem{htt} Z. Hasciewicz, K. Thielemans and W. Troost, J.~Math.~Phys.
{\bf 31} (1990) 744.
\bibitem{kn} V. G. Knizhnik, Theor.~Math.~Phys. {\bf 66} (1986) 68.
\bibitem{be} M. Bershadsky, \pl{174}{86}{285}.
\bibitem{estps} F. Englert, A. Sevrin, W. Troost, A. van Proeyen and
P. Spindel, J.~Math.~Phys. {\bf 29} (1988) 281.
\bibitem{nb} N. Berkovits, \np{358}{91}{169}.
\bibitem{bcp} L. Brink, M. Cederwall and C. Preitschopf, \pl{311}{93}{76}.
\bibitem{cp} M. Cederwall and C. R. Preitschopf, \cmp{167}{95}{373}.
\bibitem{sp} J. A. H. Samtleben, \np{453}{95}{429}.
\bibitem{nw} H. Nicolai and N. P. Warner, \cmp{125}{89}{369}.
\bibitem{bsn} E. Bergshoeff, E. Sezgin and H. Nishino, \pl{186}{87}{167}.
\bibitem{wtn} B. de Wit, A. K. Tollst\'en and H. Nicolai, \np{392}{93}{3}.
\bibitem{fl} E. S. Fradkin and V. Yu. Linetsky, \pl{275}{92}345;
             \ibid{282}{92}{352}.
\bibitem{bo} P. Bowcock, \np{381}{92}{415}.
\bibitem{imp} K. Ito, J. O. Madsen and J. L. Petersen, \pl{318}{93}{315};\\
\np{398}{93}{425}.
\bibitem{kac} V. G. Ka\v{c}, {\it Infinite Dimensional Lie Algebras.
An Introduction}, Boston: \\ Birkh\"auser, 1983.
\bibitem{fk} P. G. O. Freund and I. Kaplansky, J.~Math.~Phys. {\bf 17} (1976)
 228.
\bibitem{snr} M. Scheunert, W. Nahm and  V. Rittenberg, J.~Math.~Phys. {\bf 17}
 (1976) 1626.
\bibitem{dvn} B. de Witt and P. van Nieuwenhuizen, J.~Math.~Phys.{\bf 23} 1953.
\bibitem{mg1} M. G\"unaydin, J. Math. Phys. {\bf 31} (1990) 1776; \\
{\it The Exceptional Superspace and the Quadratic Jordan Formulation of
Quantum Mechanics}, in ``Elementary
Particles and the Universe: Essays in Honor of Murray Gell-Mann'' , ed.
by J. H. Schwarz, Cambridge: Cambridge Univ. Press, 1991. 
\bibitem{gs} P. Goddard and A. Schwimmer, \pl{214}{88}{209}.
\bibitem{gptv} M. G\"unaydin, J. L. Petersen, A. Taormina and A. van Proeyen,
\np{322}{89}{402}.
\bibitem{schoutens} K. Schoutens,  \np{292}{87}{150}; \ibid{295}{88}{634};
\ibid{314}{89}{519}.
\bibitem{ssvn} K. Schoutens, A. Sevrin and P. van Nieuwenhuizen,
\cmp{124}{89}{209}.
\bibitem{es} Z. Khviengia and E. Sezgin, \pl{326}{94}{243}.
\bibitem{ke3} S. V. Ketov, \cqg{12}{95}{925}; \mpl{10}{95}{79}.
\bibitem{gko} P. Goddard, A. Kent and D. Olive, \cmp{103}{86}{105}.
\bibitem{ks} Y. Kazama and H. Suzuki, \pl{216}{89}{112};
\np{321}{89}{232}.
\bibitem{vp} A. van Proeyen,  \cqg{6}{89}{1501}.
\bibitem{st} A. Sevrin and G. Theodoridis, \np{332}{90}{380}.
\bibitem{mg93} M. G\"unaydin and S. Hyun, \np{373}{92}{688}; \\
 M. G\"unaydin, \pr{47}{93}{3600}.
\bibitem{gk} S. J. Gates Jr., and S. V. Ketov, \pr{52}{95}{2278}.
\bibitem{gg} M. G\"unaydin and F. G\"ursey, J.~Math.~Phys. {\bf 14} (1973)
1651; \\
\bibitem{cdfn} E. Corrigan, C. Devchand, D. B. Fairlie and J. Nuyts,
\np{214}{83}{452}.
\bibitem{wn} B. de Wit and H. Nicolai, \np{231}{84}{506}.
\bibitem{dgt} R. D\"{u}ndarer, F. G\"{u}rsey and C. Tze, \np{266}{86}{440}. 
\bibitem{gn} M. G\"{u}naydin and H. Nicolai, \pl{351}{95}{169}.
\bibitem{wznw} J. Wess and B. Zumino, \pl{37}{71}{95};\\
S. Novikov, Sov.~Math.~Dokl. {\bf 24} (1981) 222, Usp.~Mat.~Nauk {\bf 37}
(1982) 3; \\
E. Witten, \cmp{92}{84}{455}. 
\bibitem{vkpr} P. Di Vecchia, V. G. Knizhnik, J. L. Petersen and P. Rossi,
\np{253}{85}{701}.
\bibitem{m1} P. Mathieu, \pl{218}{89}{185}.
\bibitem{hs} J. Harvey and A. Strominger, \prl{66}{91}{549}.
\bibitem{i} T. A. Ivanova, \pl{315}{93}{277}.
\bibitem{fny} D. B. Fairlie and J. Nuyts, {J.~Phys.} {\bf A17} (1984) 431.
\bibitem{dwn} S. Fubini and H. Nicolai, \pl{155}{85}{369}.
\bibitem{ip} T. A. Ivanova and A. D. Popov, Lett. Math. Phys. {\bf 24} (1992)
85.
\bibitem{gnst} M. G\"{u}naydin, B. E. W. Nilsson, G. Sierra and P. K. Townsend
\pl{176}{86}{45}.
\bibitem{dnp} M. J. Duff, B. E. W. Nilsson and C. N. Pope, Phys. Rep. {\bf 130}
(1986) 1.
\bibitem{dn87} B. de Wit and H. Nicolai, \np{281}{87}{211}.
\bibitem{t1} P. K. Townsend, \pl{350}{95}{184}. 
\bibitem{w1} E. Witten, \np{443}{95}{85}; {\it Some comments on string 
dynamics}, Princeton preprint IASSNS--HEP--95--63, July 1995, hep-th/9507121.
\bibitem{j} D. D. Joyce, {\it Compact  Riemannian 7-manifolds with holonomy 
$G_2$}, Oxford preprints I and II, 1994; {\it Compact Riemannian 8-manifolds 
with exceptional holonomy $Spin(7)$}, Oxford preprint, 1994.
\bibitem{sv} S. Shatashvili and C. Vafa, {\it Superstrings and manifolds with
exceptional holonomy}, Harvard and Princeton preprint, HUTP-94/A016 and
IASSNS-HEP-94/47, 1994, hep-th/9407025.
\bibitem{pt11} G. Papadopoulos and P. K. Townsend, \pl{357}{95}{300}.
 
\end{thebibliography}
\end{document}